\documentclass[a4paper,fleqn,usenatbib]{mnras}
\usepackage{mathptmx}

\usepackage[T1]{fontenc}
\usepackage{ae,aecompl}

\usepackage{pdflscape}

\usepackage{graphicx}	
\usepackage{amsmath}	
\usepackage{amssymb}	
\usepackage{ulem}           

\newcommand{\beq}	{\begin{equation}}
\newcommand{\eeq}	{\end{equation}}
\newcommand{\beqa}{\begin{eqnarray}}
\newcommand{\eeqa}{\end{eqnarray}}
 
\newcommand{\dis}{\displaystyle}
\newcommand{\e}	{$^{-1}$}
\newcommand{\ee}	{$^{-2}$}

\newcommand{\caln}	{{\cal N}}
\newcommand{\calr}	{{\cal R}}

\def\dis{\displaystyle}

\def\ltsimeq{\,\raise 0.3 ex\hbox{$ < $}\kern -0.75 em
 \lower 0.7 ex\hbox{$\sim$}\,}
\def\gtsimeq{\,\raise 0.3 ex\hbox{$ > $}\kern -0.75 em
 \lower 0.7 ex\hbox{$\sim$}\,}

\let\ga=\gtsimeq
\let\la=\ltsimeq

\def\eff	{{{\rm eff}}}

\def\M	{{\rm M}}

\def\ms    {{\rm MS}}
\def\MS    {{\rm MS}}

\def\mch	{m_{\rm ch}}

\def\msun{$M_\odot \,$}

\def\sl			{{s,\,\ell}}

\def\mch{m_{\rm ch}}

\def\eff                 {{\rm eff}}

\def\ms                  {{\rm MS}}

\def\s0{{\dot {N}} _{a0}}

\def\L0{L_{0,\rm band}}

\def\t0{t_{\rm band}}

\def\edNst		 {\dot\caln_{*T}}

\def\eNsm{\caln_*(m)}

\def\edNsmt{\dot\caln_*(m,t)}

\def\ABS		{{\rm ABS}}
\def\app		{{\rm app}}
\def\Gyr		{{\rm Gyr}}
\def\r15336      {\caln_*(1.5<m<3)/\caln_*(3<m<6)} 
\def\gap		{\Gamma_\app}
\def\gapz		{\Gamma_{\app,0}}
\def\gapo		{\Gamma_{\app,1}}
\def\gfap		{\Gamma_{f,\app}}
\def\gapot	{\Gamma_{\app,12}}
\def\msun           {$M_\odot \,$}
\def\MS		{{\rm MS}}
\def\Nref           N_{ref}
\def\obs		{{\rm obs}}
\def\SFR		{{\rm SFR}}
\def\tsfr		{{t_{\rm SFR}}}

\title[Initial Mass Function]{The High Mass Slope of the IMF}

\author[Parravano, Hollenbach, \& McKee]{
Antonio Parravano$^{1,2}$\thanks{E-mail: parravan3@gmail.com}, 
David  Hollenbach$^{3}$
Christopher F. McKee$^{4}$
\\
$^{1}${Universidad de Los Andes, Centro De F\'{\i}sica Fundamental, M\'erida 5101a, Venezuela}\\
$^{2}${Universidad de M\'{a}laga, M\'{a}laga, Spain}\\
$^{3}${SETI Institute, 189 Bernardo Avenue, Mountain View, CA 94043}\\
$^{4}${Physics Department and Astronomy Department, University of California at Berkeley, Berkeley, CA 94720}
}

\date{Accepted XXX. Received YYY; in original form ZZZ}

\pubyear{2018}

\begin{document}
\label{firstpage}
\pagerange{\pageref{firstpage}--\pageref{lastpage}}
\maketitle

\begin{abstract}
Recent papers have found that the inferred slope of the high-mass ($>1.5$ M$_\odot$) IMF for 
field stars in the solar vicinity has a larger value ($\sim 1.7-2.1$) than the slopes ($\sim 1.2-1.7$; 
Salpeter= 1.35) inferred from numerous studies of young clusters. We attempt to reconcile this 
apparent contradiction. Stars mostly form in Giant Molecular Clouds, and the more massive stars ($\ga 3$ 
M$_\odot$) may have insufficient time before their deaths  to uniformly populate the solar circle of the Galaxy. 
We examine the effect of small sample volumes on the {\it apparent} slope, 
$\Gamma_\app$, of the high-mass IMF by modeling the present day mass function (PDMF) over the mass 
range $1.5-6$ M$_\odot$. Depending on the location of the observer along the solar circle and the 
size of the sample volume, the apparent slope of the IMF can show a wide variance, with typical values 
steeper than the underlying universal value $\Gamma$.  We show, for example,  that the PDMFs 
observed in a small (radius $\sim 200$ pc) volume randomly placed at the solar circle have a $\sim 15-30$\%  
likelihood of resulting in $\Gamma_{\app} \ga \Gamma+ 0.35$  because of inhomogeneities in the 
surface densities of more massive stars.   If we add the a priori knowledge that the Sun currently lies 
in an interarm region, where the star formation rate is lower than the average at the solar 
circle,   we find an even higher likelihood    ($\sim 50-60\%$ ) of  $\Gamma_{\app} \ga \Gamma+0.35$, 
corresponding to $\Gamma_{\app} \ga 1.7$ when the underlying $\Gamma= 1.35$.

\end{abstract}


\section{Introduction}

The stellar initial mass function (IMF) is of fundamental importance in astrophysics because it relates the relative number
of high-mass stars that produce most of the light and energy to the low-mass stars that make up most of the stellar mass.   In distant and/or dust-obscured
regions of the Galaxy and in external galaxies, it is often the case that only the high-mass stellar component
can be measured, whether through direct emission from the massive stars, from 
emission from the HII regions surrounding these stars, or through reprocessed infrared radiation.
If the IMF is known, such observations can be used to estimate the total star formation rate (SFR).  
More generally, as \cite{wei15} have pointed out, knowing the {IMF of stars ``is essential to interpreting the stellar
populations of star-forming galaxies across cosmic time, testing
and validating theories of star formation, constraining models of chemical
evolution, the formation of compact objects, and understanding the interplay between
stars and gas."}

There is considerable evidence that the IMF in disk galaxies is universal
(\citealp{ren05}, \citealp{bas10}, \citealp{off14}); 
in particular, the last authors conclude that
studies of resolved stellar populations provide 
no evidence for variations of the IMF within local galaxies that are greater than 2$\sigma$.
For stellar masses $m\ga1$ (in this paper, all stellar masses are in units of the solar mass), the
IMF { has been usually modeled as} a power law, $\psi(m)\propto m^{-\Gamma}$ 
up to a limiting mass $\sim 100$,
where $\Gamma$ is the underlying universal slope of
the high-mass IMF \citep{bas10}.
These reviews conclude that the observations are consistent with $\Gamma\simeq 1.35$, the Salpeter
value. 
The value of this power law is particularly important, since it regulates the relative number
of low- and high- mass stars.
{ One reason the underlying IMF may be universal is that current theories of the origin of the 
IMF are based on turbulence, so the slope is independent of the density, chemistry and other properties. It depends primarily
on the exponent $p$ in the linewidth-size relation $\sigma \propto r^p$ (e.g., \citealp{hop12}; for supersonic turbulence, $p\simeq 0.5$} (e.g., \citealp{hen12}).

On the other hand, there is observational evidence for deviations from a universal IMF in extreme environments:
\cite{ludo13} find a smaller value of $\Gamma$ in the Galactic Center clusters, as do \citet{schn18} in 30 Dor.
A number of workers have found evidence for a bottom heavy IMF in giant ellipticals 
(e.g., \citealp{vandok10}, \citealp{cap12}). We note that
it has been suggested that the IMF on galactic scales is the superposition of
cluster IMFs that obey a relation between the maximum stellar mass and the
the mass of the cluster (e.g., \citealp{kro03}, \citealp{kro13} and references therein). Evidence against
this hypothesis and in favor of the hypothesis that cluster IMFs are the result
of random sampling from a universal IMF is summarized in \cite{ash17}. 
 In this paper we adopt the latter hypothesis, at least in the Milky Way
outside its nucleus, and we examine reasons why the observed ratio of high mass to low mass stars might vary from region
to region.

 \subsection{IMF from Clusters and IMF from Local Field Stars}

Observations of stellar clusters (including stellar associations) that are young enough
that none of the stars 
within a specified mass range 
have evolved off the main sequence allow a direct estimate of $\Gamma$, and we define
the slope inferred from such observations as $\Gamma_c$.
The primary advantage of using cluster studies 
over field-star studies to infer the slope of the IMF
 is that the stars in clusters were
all born at approximately the same time and have the same distance and
metallicity. 
Even with this simplification, however, a number of
problems remain with inferring the high-mass slope from cluster studies (\citealp{sca05}):
(i) Most clusters studied have a small number of
stars per mass bin or a very small number of bins; (ii) the ubiquity of mass segregation means that clusters must
be observed to large radii; (iii) 
membership is a problem, especially at large radii; (iv) the correction for unresolved binaries is uncertain;  (v) adopted cluster 
properties (distance, age, metallicity, extinction, and differential reddening) can introduce errors;
and (vi) uncertainties in evolutionary tracks and isochrones affect mass determinations.
Therefore it is difficult to determine if the large variation in the inferred values of $\Gamma_c$ from cluster to cluster
is due to variations in the IMF or to the uncertainties introduced by these problems.

\cite{wei15} reviewed 
the cluster  literature on the high-mass slope and concluded that most of the studies found
$1.2 \la \Gamma_c \la1.7$. 
By studying a large number (85) of clusters in M31, they were able to eliminate almost all 
the problems listed above.
They concluded that the intrinsic high-mass ($m>2$) slope there is $\Gamma_c=1.45^{+0.03}_{-0.06}$ and
that this high-mass slope for the 85 clusters is universal.

To determine the IMF from observations of field stars, observers/modelers 
match synthetic populations to counts of stars in the color-magnitude diagram.
These population synthesis simulations assume evolutionary tracks, stellar atmosphere models, a
given time evolution of the SFR (the star formation history, or SFH), vertical diffusion, binary frequency as a function of mass,  and a model for interstellar extinction.
The simulations
are run with various underlying values of $\Gamma$, and 
the best fit to the observations provides the 
apparent slope of the field star IMF,  $\Gamma_{f,\app}$. 
Here the subscript ``app" indicates that this is the apparent value one would obtain after attempting to allow for
all the uncertainties just enumerated, but assuming that stars are distributed uniformly 
in the plane of the Galaxy.   In fact, however, stars are not uniformly distributed in the plane because
star formation is highly inhomogeneous in both space and time, and the focus of this
paper is on 
the effects that this non-uniform distribution of stars, particularly the more
massive and therefore younger ones, have on the inferred high-mass slope of the IMF.

 \subsection{Three Recent Determinations of the High-Mass 
Slope of the
IMF from the Local Field }

In contrast to the determinations of $\Gamma_c \simeq 1.2-1,7$  in the Milky Way and other galaxies, there have been
three recent determinations of the apparent slope of the field-star IMF
in the solar neighborhood that have yielded higher values, $\Gamma_{f,\app} \simeq 1.7- 2.1$ (\citealp{daw10}, 
\citealp{cze14}, and \citealp{ryb15}).  
These papers assume a smooth distribution of the SFR
 in the plane of the Galaxy and  attempt to account for binarity, 
 the SFH and the age-dependent vertical distribution of stars;
the radial gradient of the SFR is neglected in these studies due to the small size of the observational volumes.

Dawson \& Schr\"oder (2010) used Hipparchos data to
create a sample of stars with $m>0.9$, within a radius of 100 pc of the Sun, and within $\pm 25$ pc of the Galactic plane. They
compared synthetic simulations that included the effect of unresolved binaries and the effect of
age-dependent vertical diffusion of stars.
For their
favored binary model, in which the binary fraction (including multiples) is 71\%,
they found
$\Gamma_{f,\app} \simeq 1.85 \pm 0.15$. If they allowed the binary fraction to decrease with mass from 71\% for 
primary stars just above  $3\,M_\odot$ to 57\% for primary stars just above $1\,M_\odot$, they found
$\Gamma_{f,\app} \simeq 2.2$, but this binary model was not favored by their data.  They concluded that the
SFR has been ``reasonably" constant during the lifetime of the stars in their sample, but they did not quantify how much
it could have changed over the last 3 Gyr, the time period which we focus on in this paper (roughly the lifetime of a $m=1.5$ star).   
One problem with inferring SFH was the uncertainty in the rate of vertical diffusion of stars out of their observing volume as
a function of stellar age.
 
 Czekaj et al (2014) compared 
 synthetic distributions of the number of stars plotted against B-V with Hipparchos (Tycho-2) data and estimated that
 $\Gamma_{f,\app} \sim 2$ in the solar neighborhood.   In contrast to \citet{daw10} and \citet{ryb15}, they considered stars over a wide mass range, although the main fitting of the model to data was 
to Tycho-2 data, which 
 is a magnitude limited sample down to $V=11$.
They used an updated version of the Besancon Galaxy model and tested
 several different SFHs, IMFs, evolutionary tracks, and dust extinction models against the data.   
 They corrected for the effect of binaries by using the binary fraction determined by \citet{are11}, which varies from about
 0.1 for $m=0.1$ to 0.84 at very high masses.   The SFHs include both a constant SFR and a rate that declines as 
 $\exp(-t/t_{\SFR})$, where
 $t_{\SFR}=8.3$ Gyr (\citealp{aum09}).\footnote{Note that this is equivalent to the form $\exp(-0.12t_{\rm Gyr}$), where $t_{\rm Gyr}$
 is $t/1$ Gyr.   This form is often seen in the literature.}   They focused on stars
 with $V<11$, which means that the volume included around the Sun depends on the stellar luminosity as well as
 the dust extinction corrections; stars of mass $m\sim 1.5$ are included in their sample provided they are within about 300 pc
 for a typical amount of extinction.
 There is no error estimate possible in this study because it included only 5 different values of $\Gamma$ in the simulations:
two simulations with $\Gamma = 1.3$, six with $\Gamma=2.0$, one with $\Gamma=2.2$, one with $\Gamma=2.5$, and one with $\Gamma=3.16$.  The chi squared fits
 favor $\Gamma =2$ for these simulations; the $\Gamma= 1.3$ and $\Gamma= 3.16$ cases have values of chi squared 
 that are about
 3 times larger than the favored case.  However, there are no simulations with $\Gamma$ between 1.3 and 2, so we cannot know
 where the minimum chi squared occurs.  The authors concluded that $\Gamma_{f,\app}$ is larger than 1.3 and is about 2.
 
 Analyzing stars within 200 pc from Hipparcos and from the Catalogue of Nearby Stars,
 \cite{ryb15} improved upon an earlier 
paper by \citet{jus10} and found that a better treatment of
binarity and reddening reduced $\Gamma_{f,\app}$ from 3.16 in \citet{jus10} to $ 2.02 \pm 0.06$ for the mass range $1.4<m<10$.   
This demonstrates the sensitivity of the inferred value of $\Gamma_{f,\app}$ 
 to these corrections. Their adopted SFH
 scales as exp$(-t/t_{\SFR})$ \citep{aum09} for the last several Gyr (again with $t_{\SFR}=8.3$ Gyr). They also 
 allowed for vertical diffusion of stars with age.  
 The quoted errors are very small, but as the authors note, these errors do not include systematic errors due
 to uncertainties in the binary correction, the inhomogeneous ISM, variable stars, the thick disk component, the metal enrichment law,  and low mass stellar envelopes.
 
 Given the discussion above, we take these three papers in aggregate to indicate that the apparent value of the
 high-mass slope of the IMF, after correction for a possible decline in the SFR with time and for vertical diffusion of stars due to disk heating, is $\Gamma_{f,\app} \simeq 1.7-2.1$ 
 in the solar neighborhood.    We note the sample volumes in \cite{daw10} and \cite{ryb15} are
 relatively small, with radii $<$ 200 pc.

 \subsection{Summary of Goals of the Paper and the Methodology of  our Simulations}
 
 In this paper we explore the possibility that the apparent discrepancy between the estimates from young clusters
 and galaxies discussed in the above reviews, which have $\Gamma_c \sim 1.2-1.7$, and the local estimates 
 of $\Gamma_{f,\app}  \simeq 1.7-2.1$ can be explained by 
  clustered star formation and by the location of the Sun in an interarm region.   Our simulations isolate the effect of
  the inhomogeneous distribution of 
  higher mass ($m\ga 3$) stars on the inference of the high-mass IMF slope when
  observing finite volumes at the solar circle.  
  
  Massive stars have an inhomogeneous distribution in the Galactic plane
  because they are mostly born in massive GMCs that sparsely populate the Galaxy, and their short lifetimes do not allow
  them to move from these localized birthsites to uniformly populate the plane.   
 Lower mass stars live longer, so they diffuse from their birth sites and spread more uniformly in the plane.
 Thus, there will be substantial
 volumes of Galactic space around the solar circle with a relative deficit of higher mass stars (high $\Gamma_{f,\app})$, and at the same time smaller volumes that include
 significant sites of recent star formation with relatively high numbers of higher mass stars (low $\Gamma_{f,\app})$. 
  Furthermore, the Sun is not randomly placed along the solar circle, but in an interarm region,  
 where the SFR has been depressed compared to the solar circle average for $\sim 100$ Myr.
 As \citet{elm06} have pointed out, this has most likely steepened the PDMF and could lead to the erroneous inference that the high-mass slope of the IMF is steeper than it actually is. 
 
 The simulations we present in this paper can be summarized as follows (details in Section 3).
 Our simulation volume
includes a large portion of the solar circle.   Stars are born with an IMF characterized by an underlying $\Gamma$ 
inside widely separated GMCs,
and they then move from their birthsites
to fill the volume of the simulation.  
Their velocities are prescribed from observations of stellar velocity dispersions in the plane of the Galaxy.
 Given these velocities, there is not enough time during their lifetimes for the higher mass stars to homogenize; the higher the mass, the shorter the 
 lifetime and therefore the greater the fluctuation in density.
 We follow stars of mass
 $m= 1.5-6$ because $m\ga 1.5$ ensures we are on the high mass slope part of the IMF, and because
 a range of mass is needed to provide an accurate measure of slope.\footnote{Because of the steepness of the PDMF,
 our measure of the slope depends mostly on the ratio of $m=3$ stars to $m=1.5$ stars, so that the results do not
 depend sensitively on the upper limit value $m=6$ (as will be discussed further in Section 2).}
  In order to make our procedure more closely parallel what the observers do, we include the effects of a time 
  dependent SFR and an age-dependent vertical diffusion in our
simulation and then remove these effects afterwards in the same manner that observers would.    We run the simulation for 3 Gyr so that we get a full
sample of all MS stars with mass $m>1.5$.  At the end of the simulation we know the location and mass of each of the roughly $2 \times 10^7$ stars in the simulation.   
We measure the PDMFs in random small (radius $\sim 100-750$ pc) volumes at the solar circle
 in each simulation.  The PDMFs vary from volume to volume, and from their distribution we infer
 a distribution function of the apparent  $\gfap$
  given a fixed universal value of $\Gamma$ for stellar birth. 
  
 In contrast to the value of $\gfap$ from observations, which can differ from the underlying value of $\Gamma$ due to 
 the clustering of star formation in space and time as well as to many other errors introduced by the
 comparison of observations with the population synthesis simulations,
the apparent value of $\Gamma$  inferred from the simulations, $\gap$, differs from $\Gamma$ only because of the clustering.
For the relatively small 
volumes in the papers cited above, the PDMFs of the $m\ga1.5$ stars can differ significantly from
that due to a uniform distribution of stars, resulting in
 a significant probability of finding $\Gamma_\app > \Gamma$ from a randomly chosen observational location at the solar circle.
 The effect is larger if one considers that the Sun is not in a random position but in an interarm position.
 
 The purpose of our simulations is not to compare simulation with observation directly, but rather to
demonstrate the effect that the inhomogeneous distribution of stars can have in estimating $\Gamma$. 
Our simulations provide a quantitative estimate of the magnitude of this single effect.

   Section 2  presents the basic theory that relates the IMF to the PDMF, and a simple analytic fit to
 the stellar lifetimes that enables an analytic solution to the  PDMF.
Section 3  presents our model for the spatial  distribution of stars of various masses at
 the solar circle, given that stars are born in GMCs and diffuse from these formation sites. These
 numerical simulations are repeated numerous times, and from these simulations we determine
 the distribution of $\Gamma_\app$ as a function of the volume of space sampled in 
 observationally determining the PDMF. 
 Section 4 focusses on the effect of spiral arms, providing  representative examples of the effects that spiral
 arms could have on the distribution of $\Gamma_{\app}$.
 Finally, in Section 5 we present a summary and our conclusions, and note the relevance of these
 results to the Gaia data.

\section{Inferring the High-Mass IMF from the PDMF}

 In this section we derive analytical expressions that relate the PDMF to the IMF.
Although the high-mass portion of the IMF is well characterized by a power law (\citealp{bas10}), the high-mass
PDMF is not necessarily so described. We therefore consider a range of masses,
$m=$1.5 to 6, chosen to be well represented in the observational studies
of \citet{daw10}, \citet{cze14} and \citet{ryb15} and to lie clearly in the high mass regime of the IMF.
We characterize the PDMF here and in our simulations by the ratio $\calr$ of lower mass 
($1.5 < m < 3$) stars to higher mass ($3 < m < 6$) stars in a given finite volume:
\beq
\calr\equiv {\caln_*(1.5<m<3) \over \caln_*(3<m<6)}  .
\label{eq:calr}
\eeq
Due to the steepness of the PDMF, this ratio is essentially the ratio of stars with mass $m\sim 1.5$ to
 stars with mass $m \sim 3$.  Under the assumption of a homogeneous distribution of stars of all masses in
 the plane of the Galaxy, we analytically derive below  how the relation of $\calr$ to the underlying $\Gamma$
 varies due to the SFH (constant SFR versus declining SFR) and age-dependent vertical diffusion.   The latter
 effect depends on the volume sampled and we show this dependence.   In Section 3 we treat the
 effect of the inhomogeneous distribution of stars of various masses in the plane of the Galaxy due to their
 localized (in time and space) births.  This effect is not amenable to an analytic treatment and so is not
 included in this section.
 
 We emphasize that we use the ratio $\calr$ to characterize the PDMF and to infer the IMF in our simulations.  The
 papers that compare population synthesis models with observations do not use this ratio.   However, in essence, they are comparing
 the densities of higher mass and lower mass stars in a given volume to infer the slope of the IMF, so their method
 is analogous to ours.   The point of the simulations is not to completely mimic the population synthesis methods, but to isolate
 the effect that the localized star formation has on the inference of the IMFs from observed PDMFs in small volumes.
 An advantage the simulations is that we know the masses of the stars and their locations, so there are no errors 
 caused by model atmospheres, distance ambiguities, binary corrections, observational errors, etc.

\label{sec:theory}

\subsection{The IMF and the PDMF}

We begin by relating the high-mass portion of the IMF to the total IMF.
Let $\psi(m) d\ln m$ be the fraction of stellar objects (including brown
dwarfs) born in the mass range $m$ to $m+dm$.
In Parravano, McKee \& Hollenbach (2011, henceforth Paper I),
we adopted the functional form for the IMF,
\beq
\psi(m)=C m^{-\Gamma}\{1-\exp[-(m/\mch)^{\gamma+\Gamma}]\},
\label{eq:imf}
\eeq
where $C$ is a normalization constant such that
\beq
\int_{m_\ell}^{m_u}\psi(m) d\ln \,m =1.
\eeq
 This form for the IMF
approaches a power law both at low stellar masses,
\beq
\psi\rightarrow C\mch^{-\Gamma}\left(\frac{m}{\mch}\right)^{\gamma},
\eeq
and at high stellar masses,
\beq
\psi\rightarrow C m^{-\Gamma}, 
\label{eq:imfhi}
\eeq
and we termed it
the Smoothed Two-Power Law (STPL) form for the IMF. 
 Assuming $\Gamma=1.35$, we determined best fit values of $\gamma=0.51$ and $\mch=0.35$,
corresponding to $C=0.108$, for the solar neighborhood. Insofar as the IMF
is universal, as discussed in Section 1, this IMF should be generally valid in disk galaxies.

We note that other authors have adopted different functional forms for the IMF, but
there is general agreement that for $m\ga 1$, the IMF follows a power law $\psi(m) \propto  m^{-\Gamma}$
 (\citealp{bas10}).
Because this paper treats stars with $m>1.5$, we adopt here the high-mass form 
from Equation (\ref{eq:imfhi}), $\psi(m)=Cm^{-\Gamma}$ (this form agrees with Equation 2 to better than 0.1\% for
$m>1$).

 	The  differential star-formation rate by number as a function of mass and time is
given by $d\edNsmt/d\ln m$, where $\edNsmt$ includes only
stellar births, not stellar deaths and where $m$ is the initial mass of the star.
This rate can be expressed as
the product of the total 
star-formation rate by number, $\edNst(t)$, and
the probability that a star
is born with a mass $m$---i.e., the IMF, $\psi(m)$:
\beq
\frac{d\edNsmt}{d\ln m}=\edNst(t)\psi(m).
\label{eq:separable}
\eeq

The PDMF can be expressed as $d\caln_*(m)/d\ln m$, where
$d\eNsm$ is the number of 
stars with masses between $m$ and $m+dm$
at the present epoch, $t=t_0$.
This form of the PDMF is 
about equal to the number of stars in a mass interval of a factor $e$ around $m$.  
To relate the PDMF to the IMF, we make two assumptions: First, we assume that
the stellar mass is about constant in time as the star ages--i.e., we focus on main sequence stars that are not
too massive. Second, we assume that the volume in which the PDMF
is measured includes all the surviving main sequence stars that were
formed. Then the PDMF is the time integral of the 
star formation rate,
\beq
\frac{d\caln_*(m)}{d\ln m} \ =\ \int_{t_0-\tau'(m)}^{t_0} \edNst(t)\psi(m)\; dt,
\label{eq:pdmf1}
\eeq
where $\tau'\equiv \min[\tau_\ms(m),\,t_0]$ and $\tau_\ms(m)$ is the main sequence lifetime of
a star of mass $m$  (e.g., \citealp{sca86} ). For the particular case of stars with $m\ga 1$, so that $\tau_\ms(m)<t_0$, and
for a constant star formation rate, the PDMF is simply
\beq
\frac{d\caln_*(m)}{d\ln m}  \ =\ \edNst\psi(m)\tau_\MS(m).
\label{eq:pdmf2}
\eeq
 We note a similar derivation in the \cite{pra08} review (Eqs. 2.20, 2.2.1).

\subsection{Relation of $\calr$ to IMF Slope for Constant SFR}
\label{sec:char}

The main effect that alters the value of $\calr$ (see Eq. 1) from its initial value is stellar evolution, since the more massive the star, the
more rapidly it evolves off the main sequence.
If the SFR has been constant for more than the lifetime of a $m=1.5$ star, $\calr$ can be derived from
Equation 8:
\beq
\calr_0=\ {{\int_{1.5}^3 \tau_\MS(m) \psi(m) d\ln m}\over 
{\int_3^6 \tau_\MS(m) \psi(m) d\ln m}},
\eeq
 where the subscript "0" refers to the assumptions of constant SFR, no vertical segregation by mass,
and homogeneous distribution in the plane of the Galaxy.
Using main sequence lifetimes from \citet{bre12}  (for example the lifetimes of stars of mass $m=1.5, 3,$ and 6 
with solar metallicity are 2570 Myr, 354 Myr and 69 Myr, respectively),
we have numerically integrated Equation (9) and found  that $\calr_0= 16.9$
for $\Gamma=1.35$. On the other hand, $\calr_0=26.6$ for $\Gamma=2$, substantially higher.
This gives an idea of how many $m=3-6$ stars would need to be depleted in order for our inferred $\Gamma_\app$ to
be 2 and not 1.35: One needs to lower $\caln_*(3<m<6)$ by [1-(16.9/26.6)]x100= 36\%.

We can obtain an approximate expression for $\calr_0$ by approximating the main sequence lifetimes of stars
in the range $1.5<m<6$,
\beq
\tau_\MS(m) \simeq 390 \left({m \over 3}\right)^{-2.72}\ {\rm Myr},
\label{eq:taums}
\eeq
which fits the lifetimes to within 9\%.
With this power law, $\tau_\MS(m)\psi(m)$ is proportional to $m^{-2.72-\Gamma}$,
 and  we can analytically compute the ratio $\calr_0$ for arbitrary $\Gamma$:
 \beq
\calr_0\simeq 2^{2.72+\Gamma}.
\eeq
 The values of $\calr_0$ obtained with this power law are accurate to within 1\% for $1< \Gamma <2$.
For example, assuming $\Gamma=1.35$ ($\Gamma=2.0$), this approximation gives $\calr_0=16.8$ ($\calr_0=26.4$) 
in excellent agreement with the exact values.

This equation can be rearranged to provide an approximation for the inferred $\gapz$ if the ratio $\calr$
is observed in some volume of the Galaxy or of our simulation:
\beq
\gapz \simeq\  1.44 \ln\calr - 2.72.
\label{eq:gapz}
\eeq
 This would provide a good estimate of $\Gamma$, the underlying slope
of the universal IMF, only if there were no vertical stellar diffusion, no time dependence of
the SFR over the last 2.6 Gyr, and no spatial variation in the SFR.
Discussion of these effects, especially the last effect,  occupies much of the rest of this paper.

\subsection{Approximation for a Declining SFR}
\label{sec:decline}

We treat the simple case of the SFR rate declining as $\edNst(t) \propto \exp(-t/\tsfr$),
with $\tsfr=8.3$ Gyr for numerical evaluation \citep{aum09}.
Define $t'\equiv t_0-t$. Equation  (6)
then gives
\beqa
\frac{d\caln_*(m)}{d\ln m}  &=& \edNst(t_0)\psi(m)\int_0^{\tau_\MS(m)} e^{t'/\tsfr} dt',\\
\label{eq:pdmf3a}
&=&\edNst(t_0)\psi(m)\tsfr \left[ e^{ \tau_\MS(m)/\tsfr} - 1\right]. 
\label{eq:pdmf3b}
\eeqa
Since we are now including only the first effect that can alter the observed value of $\calr$, a time-varying star formation rate, we label the resulting value of $\calr$ as $\calr_1$.
To solve analytically for the ratio $\calr_1$ requires that we expand the exponential as a power series in the quantity 
$\tau_\MS(m)/\tsfr$, which is small for $m=1.5-6$ stars and our adopted decay time.   This then gives the solution
\beq
\frac{d\caln_*(m)}{d\ln m}  \simeq  \edNst(t_0)\psi(m)\tau_\MS(m) \left[ 1+ { \tau_\MS(m)\over 2\tsfr}\right].
\label{eq:pdmf4}
\eeq

Evaluating $\calr$ from Equation (\ref{eq:calr}), we find
\beq
\calr_1\simeq 2^{2.72+\Gamma}\left(\frac{1+B}{1+2^{-2.72}B}\right),
\eeq
 where the subscript ``1" refers to the assumption of an exponentially declining SFR, as well as
 no vertical segregation by mass
and homogeneous distribution in the plane of the Galaxy,
and where
\beq
B\equiv \frac{1}{2 \times1.5^{2.72}}\left(\frac{2.72+\Gamma}{5.44+\Gamma}\right)\left(\frac{1-2^{-(5.44+\Gamma)}}{1-2^{-(2.72+\Gamma)}}\right)\left(\frac{7.74\,\Gyr}{\tsfr}\right).
\eeq
Since $B\sim 0.1$ is small, $\calr_1$ can be approximated as
\beq
\calr_1\simeq 2^{2.72+\Gamma+(1.2\,\Gyr)/\tsfr},
\eeq
 which is accurate to within 2.4\% for $1< \Gamma <2$.
Correspondingly, if the ratio of $m=1.5-3$ stars to $m=3-6$ stars is $\calr$, then the value of $\Gamma$ inferred
under the assumption that the only effect altering $\calr$ is an exponentially declining SFR is
\beq
\gapo \ \simeq \ 1.44 \ln\calr -2.72 -(1.20\,\Gyr)/\tsfr.
\eeq
We recover $\gapz$ for a constant SFR ($\tsfr=\infty$). Because the
notation becomes somewhat complicated, we present in Table 1 a list of parameters and their definitions.

\begin{table*}
\caption{Definitions}
\label{tab1}
\begin{tabular}{lll}
\hline
\hline 
$ \calr $             & \vline & Ratio of $m=1.5-3$ stars to $m=3-6$ stars.\\
$ \calr_0$            & \vline & Theoretical value of $\calr$ for a steady and uniform SFR.\\
$ \calr_1$            & \vline & Theoretical value of $\calr$ for a declining SFR.\\
$ \calr_{12}$         & \vline & Theoretical value of $\calr$ taking into account the effect of 
                       a declining SFR \\& \vline & and the age-dependent vertical diffusion.\\
\hline                      
$\Gamma$              & \vline & Slope (negative) of the individual IMF for over-solar stars.\\  
$\Gamma_{\app,0}$      & \vline & Value of $\Gamma_{\app}$ derived from $ \calr $ when SFR is constant and uniform.\\
$\Gamma_{\app,1}$      & \vline & Value of $\Gamma_{\app}$ derived from $ \calr $ assuming a declining SFR.\\
$\Gamma_{\app,12}$     & \vline & Value of $\Gamma_{\app}$ derived from $ \calr $ taking into account the effect
                       of a declining SFR \\& \vline & and age-dependent vertical diffusion.\\                  
$\Gamma_c$            & \vline & Value of $\Gamma$ inferred from cluster observations.\\
$\Gamma_{f,\app}$& \vline & Apparent value of $\Gamma$ inferred from the PDMF of field-star in limited volumes.  \\
$\Gamma_{\app}$ & \vline & Apparent value of $\Gamma$ from simulations, generally after
	accounting for stellar \\ & \vline &evolution, a declining SFR, and vertical diffusion
	so that $\Gamma_{\app}\equiv \Gamma_{\app,12}$.\\
$\Delta \Gamma_{\app}$ & \vline & Difference between $\Gamma_{\app}$ and the underlying $\Gamma$.\\
\hline
\end{tabular}
\end{table*}

For $\tsfr=8.3$ Gyr and  $\Gamma=1.35$, we have $\calr_1\simeq18.6$.   Had we assumed a
constant SFR (as \citealp{daw10} did) and had we used this ratio to determine $\Gamma_\app$ from Equation (\ref{eq:gapz}), then we would have
obtained $\Gamma_\app= 1.49$ instead of the underlying $\Gamma=1.35$.   Hence, a declining SFR with
$t_{\SFR}=8.3$ Gyr can account
for only a part of the discrepancy between
the  underlying $\Gamma$  and the value $\Gamma_\app \simeq 1.85$ derived from
limited volumes around the Sun by \cite{daw10}.   We can use Eqs (11) and (18) to derive
that $t_{\SFR}= 2.4$ Gyr is needed to produce $\Gamma_{\app,0}=1.85$ if the underlying $\Gamma= 1.35$ and the star formation
is declining as given by $t_{\SFR}$.   Such a rapidly decreasing SFR has not been inferred in studies of the SFH over the past
$\sim 3$ Gyr.   Similarly, although the studies of \cite{cze14} and \cite{ryb15} assumed $t_{\SFR}= 8.3$ Gyr, the
true $t_{\SFR}$ would have to be much smaller in order for their results $\Gamma_{f,\app} \sim 2$ to be consistent with
an underlying $\Gamma \sim 1.2-1.7$.  Such a small $t_{\SFR}$ seems unlikely.  Instead,
the explanation for why the recent field star results find $\Gamma_{f,\app}  \sim 1.7-2.1$ compared to what could be a much
lower underlying $\Gamma$ may lie in the effect of clustered
star formation, which we treat below in Section 3.  However, first we treat the effect of the age dependent, and
therefore mass dependent, vertical distribution of stars, which can affect the derivation of $\Gamma$ from limited volume
samples.

\subsection{Effect of Age-dependent Vertical Diffusion}
\label{sec:vert}

Since stars of different ages have different vertical scale heights, the observed value of $\calr$ depends 
on the volume within which the star counts are made. If the observed volume is a sphere of radius $r_\obs$ centered at the mid plane,
then the ratio of stars of age $t^\prime$ that are within the sphere to the stars of the same age in a cylinder of the same radius but infinite $z$ height is
\beqa
F(t^\prime,r_\obs)=& \dis\frac{\dis\int_0^{r_\obs}\left[1-\left(\dis\frac{z}{r_\obs}\right)^2\right]
\exp\left[-\dis\frac{z}{h(t^\prime)}\right] dz}
{\dis\int_0^\infty \exp\left[-\dis\frac{z}{h(t^\prime)}\right] dz},\\ 
=& 1-2\left[\frac{h(t^\prime)}{r_\obs}\right]^2 + \frac{2 h(t^\prime)}{r_\obs}\left[1+\frac{h(t^\prime)}{r_\obs}\right]
\exp\left[-\frac{r_\obs}{h(t^\prime)}\right].
\label{eq:f}
\eeqa
Since the velocity dispersion of the stars increases with time, their scale height increases with age. We adopt
\beq
h(t^\prime) = \left\{ \begin{array}{ll}
\sqrt{45^2 +(t^\prime/0.0024\ {\rm Gyr} )^2} {\rm pc}  & ~~~~{\rm if} \,\, 0\leq t^\prime \leq 0.5 {\rm Gyr} \\
177 (1+t^\prime)^{0.5} {\rm pc}  & ~~~~{\rm if} \,\,  t^\prime > 0.5 {\rm Gyr} \\
\end{array}
\right.
\label{eq:htSD03}
\eeq
 the scale height for thin disk stars from \cite{schr03}.

Allowing for both (1) an exponentially declining SFR and (2) vertical diffusion, the ratio $\calr_{12}$ is given as
 \beq
\calr_{12}(r_\obs,\tsfr, \Gamma)=\frac{\dis\int_{1.5}^{3}m^{-\Gamma}\left[\int_0^{\tau_\MS(m)} F(t^\prime,r_\obs) e^{t^\prime/\tsfr} dt^\prime
 \right]d\ln m}
{\dis\int_{3}^{6}m^{-\Gamma}\left[\int_0^{\tau_\MS(m)} F(t^\prime,r_\obs) e^{t^\prime/\tsfr} dt^\prime \right]d\ln m}.
\label{eq:r}
\eeq
When $r_\obs \rightarrow \infty$, we see from Equation (\ref{eq:f}) that $F(t,r_\obs) \rightarrow 1$, as expected,
so that the values of $\calr_{12}$ given by Equation (\ref{eq:r}) are 
the same as
the values derived previously in Equation (\ref{eq:pdmf3a}), where we assumed the stars of all masses well mixed.  For example,
with infinite $r _\obs$ and with $\Gamma=1.35$, the $\calr$ values for $\tsfr=\infty$ and $\tsfr=8.3$ Gyr are respectively
$\calr=16.9$ and $\calr=18.6$. For finite values of $r_\obs$ the values of $\calr$ 
are reduced since $F(t,r_\obs)$ is smaller for the 
older (and therefore preferentially lower mass) stars with larger scale 
heights, fewer of which are captured in the observational volume.

 \begin{figure}
\hbox{\includegraphics[width=\columnwidth]{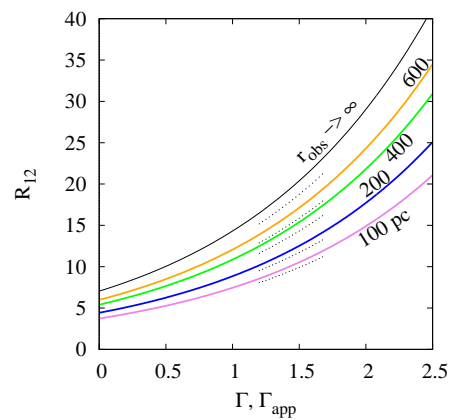}}
\caption{The ratio $\calr_{12}\equiv\r15336 $ within a sphere of radius $r_\obs$ centered at the mid-plane as a function of the 
underlying IMF slope $\Gamma$ when 
the star formation is uniform in the Galactic plane, exponential in the $z$ direction,
but age (and thus mass) dependent as given by Equation (\ref{eq:htSD03}).
The assumed SFH timescale is $\tsfr=8.3$ Gyr for the solid lines and $t_{\SFR}= \infty$ (constant SFR)  for the dotted lines.
The purple, blue, green and orange curves correspond respectively to the cases with $r_\obs=100$, 200, 400 and 600 pc. The black curve corresponds to the case
$r_\obs \rightarrow \infty$ where the effect of $z$-dispersion is irrelevant.
In the latter case, $\calr_{12}$ reduces to $\calr_1$.
 }
\label{fig1} 
\end{figure}

Figure 1 shows the results from Equation (\ref{eq:r}) of varying the underlying  $\Gamma$, the observational radius $r_\obs$,
and the SFH ($\tsfr$)  on the ratio $\calr_{12}$, the value of $\calr$ when both an exponentially declining average SFR and
age-dependent vertical diffusion are taken into account.   Correspondingly, for a given observed value of $\calr$, 
the value of $\Gamma$ inferred from this figure is $\gapot$.  For simplicity, since we use this Figure 1 in 
all the results reported in the next section, we reduce the notation to
$\Gamma_{\app} \equiv \gapot$.  
 In this figure we plot two SFHs, $\tsfr=8.3$ Gyr and $\tsfr =  \infty$ (a constant SFR).   
 The ratio  $\calr_{12} $ increases with the steepness of the slope $\Gamma$,
as expected. The main thing to notice is that the ratio $\calr_{12} $ decreases as $r_\obs$ decreases:
Recall that the scale height of the stars with masses in the range 1.5-3 \msun (older stars)  is larger than of the stars in the 
mass range 3-6 \msun (younger stars). Therefore, smaller observational volumes exclude a larger number of less massive stars, 
reducing $\calr_{12}$.  Hence the effect of vertical diffusion  for finite 
$r_\obs$ is opposite to that of a declining SFR:
a shrinking $r_\obs$ decreases $\calr_{12}$, 
whereas a shrinking $\tsfr$ increases $\calr_{12}$.

There are two mechanisms which cause the observed
ratio, and therefore the inferred $\gapot$, to vary in volumes sampled.   (1) Statistical fluctuations (Poisson noise) cause
minor changes due to finite volumes and therefore finite number of stars sampled. (2) Clustered star formation causes
inhomogeneities in especially the high mass stars in the plane. 

Figure 1 does not include the effect of statistical fluctuations of the number
of stars within the observing volume. Even if we assume that star formation
is uniform in the Galactic plane, these fluctuations cause variations of the ratio
$\calr$. However, for the average rate of star formation in the solar circle, these  
variations are negligible for $r_{\obs} \ge 100$ pc since the number of stars
within the volume are large.  For example, about 1000  stars with masses in the range 3-6 \msun
lie within $r_{\obs}= 200 $ pc, leading to a statistical variation of about 0.03 in the inferred $\gapot$. 
 For $r_{\obs} = 100$ pc the standard deviation of $\calr$ due
to the clustering of the star formation process is more than four times larger than
that due to the statistical fluctuations. Therefore, although it is strictly true that
 $\gapot = \Gamma$, the underlying slope
of the universal IMF, only if 
$r_\obs\rightarrow\infty$ so that spatial variations are averaged out, the statistical
variations are sufficiently small that essentially  $\gapot \simeq \Gamma$ for all
observing radii greater than about 100 pc, {\it as long as the SFR is uniform in the Galactic plane}.

In the simulations we describe in the next section, the local SFR varies in both space and time (clustered star formation)
 so that the IMF slope $\gapot$
derived from the value of $\calr_{12}$ can be significantly 
different than the underlying $\Gamma$ for finite volumes sampling the PDMF. 
In relating $\calr_{12}$ to $\gapot$, we have accounted for the effects of a non-constant SFH 
and the vertical diffusion, but not the
effects of clustered star formation.  The complicated modeling done by observers to infer $\Gamma$
from PDMFs observed in finite local volumes of the Galaxy are quite analogous to this procedure: They are really
measuring $\gapot \equiv \Gamma_{\app}$ (assuming no observational errors and perfect assumptions on binarity, stellar atmospheres,
interstellar extinction, etc) since they do not account for variations in the SFR 
in the plane of the Galaxy.

\section{Numerical Simulations of the PDMF for Clustered Star Formation}
\label{sec:global}

 The goal of this section is to numerically model the formation of stars of mass $1.5<m<6$
in GMCs, and to follow their movement in time away from their birthsites to populate the 
the solar circle.
In our simulations we input the SFH and the vertical 
diffusion
so we can use the proper curve on Figure 1 to obtain $\Gamma_{\app}$, which differs
from $\Gamma$ 
due to spatial variations in stellar densities, particularly of the higher mass stars, 
as well as the statistics 
of small numbers of 
such
stars in small volumes.  We numerically compute the densities 
of stars of mass $1.5<m<6$, 
evaluate $\calr$ in small volumes
in the simulations, and then use Figure 1 to account for the SFH and the 
vertical segregation to derive $\Gamma_{\app}$,
the value observers would infer based on their assumption that star formation is not clustered.
With many simulations, we obtain the probability distribution of $\Gamma_{\app}$ for a given
volume sampled.   We also look at the effect of the Sun's location in an interarm region of
the solar circle in the following section.

\subsection{Model}
\label{subsec:ingrediends}

\subsubsection{Summary of Basic Model}

In order to determine the effect of spatial and temporal fluctuations in the SFR on the PDMF, 
we model a large volume around the solar circle in a frame that moves with the galactic rotation at the solar
circle.  The $x,y$ coordinates are in the plane and cover 25 and 1.5 kpc respectively, with the 25 kpc representing about one half of
the solar circle and the 1.5 kpc the radial extent of the model.  GMCs form randomly (or concentrated in spiral arms) along the solar
circle and live for 20 Myr.   The stars are born in GMCs during the GMC lifetime
and given random velocities in the plane.
We follow the motion of the  stars in the Galactic plane for 2.57 Gyr, since our least massive star considered, 
$m=1.5$, has this lifetime.  At the end of the simulation  $\sim 2.4 \times 10^7$  stars (all with masses in the range $m=1.5-6$) 
have been formed in the simulation area but only $\sim 8 \times 10^6$ 
remain on the main sequence. These main sequence stars are distributed in the simulation volume in $z$  with an age dependent scale height given by  Equation (\ref{eq:htSD03}).
We then compute ratios $\calr_{12}$ in many small volumes centered on the midplane of the Galaxy at the solar circle, and from these
many volumes, and a large number of independent simulations, use the distribution of $\calr_{12}$ to determine 
the distribution of $\Gamma_{\app}$ from Figure 1 for a 
a range of values of the
underlying $\Gamma$.   For the convenience of the reader we present here Table 2 that lists the parameters used 
in the simulations, identifies which were fixed and which were varied, and specifies the values or value range of these parameters.

\begin{table*}
\caption{Model Parameters and their Values}
\label{tab2}
\hspace{-0.5cm}\begin{tabular}{lllllll}
\hline
\hline
      & \vline & Description                              & \vline & Standard Value                &\vline & Alternative Values \\
\hline

$r_{\obs}$      & \vline & Radius of  observed volume.                & \vline & 200 pc                        &\vline &100, 400, 600, 750 pc\\
\\
        &        & {\hspace{-2 cm}\underline{{\bf Star Formation}}}                           &  &                  &  &   \\
SFR             & \vline & Star Formation Rate in $3 \leqslant R_G \leqslant11$ kpc. & \vline & 1.6 \msun/yr                  &\vline & \\
$R_{\odot}$     & \vline & Galactic radius of the solar circle.       & \vline & 8.3 kpc                       &\vline & \\
$H_{R}$& \vline & Radial scale length of stars and gas.   & \vline & 2.0 kpc                       &\vline & \\
$ \tsfr$        & \vline &  Global SFR decay time.                & \vline & 8.3 Gyr                       &\vline & $\infty$ \\
\\
       &        & {\hspace{-2 cm}\underline{{\bf Initial Mass Function of individual stars}}  }         &  &                & &  \\  
$\Gamma$        & \vline &
 $-$  Slope of the underlying high mass IMF.    & \vline & $1.2 \leqslant \Gamma  \leqslant 1.7$ &\vline & \\
$C$             & \vline & IMF normalization constant ($0.004 < m < 120$). See eq. 2& \vline & $C(\Gamma=1.35)=0.108 $&\vline & \\
\\
      & & {\hspace{-2 cm}\underline{{\bf GMC properties}}}                           &  &                  &  &   \\                                    
$\tau_{\rm GMC}$& \vline &  Lifetime of GMCs.                         & \vline & 20 Myr                        &\vline & \\
${\cal N}_{\rm cu}$ & \vline &  See eq. 24.                              & \vline & 63   &\vline & 38$^a$\\
$\gamma_c$      & \vline & $-$ Slope of the GMC mass distribution.     & \vline & 0.6                           &\vline & 0.6$^a$, M17 $^b$\\
$(M_l,M_u)$     & \vline & Lower and upper mass limits of GMCs.         & \vline &($10^5,\,6\times 10^6$) \msun   &\vline & ($10^5,10^7$) \msun $^{a,b}$ \\
\\
        &        & {\hspace{-2 cm}\underline{{\bf Stellar Dispersion}}}                          &  &                  &  &   \\
$h(\tau)$        &\vline & Vertical scale length of stars of age $\tau$.& \vline & see eq. 22       &\vline & \\
$\sigma_{\ABS}(\tau)$&\vline & In-plane velocity dispersion of  stars with ages $\tau\geqslant 500$ Myr.& \vline & see eq. A.1&\vline &\\
$\sigma_{min}$  &\vline & In-plane velocity dispersion of stars at birth.         & \vline & $6 \ {\rm km \, s}^{-1}$&\vline &\\
$\sigma(\tau)$  &\vline & In-plane velocity dispersion of  stars of age $\tau$.  & \vline & see eq. A.2    &\vline & $\sigma_{min}, \sigma_{\ABS}$\\
\\
       &        & {\hspace{-2 cm}\underline{{\bf         Properties of spiral arms}}}                           &  &                  &  &   \\
$N_S, N_W$ &\vline & number of strong and weak spiral arms.& \vline & 2, 2  &\vline & \\
$F_S, F_W$ &\vline & Fraction of GMCs formed in the strong and weak arms.& \vline & 1/2, 1/5  &\vline & 1/3, 1/6 \\
$\alpha$  &\vline & Pitch angle of spiral arms.& \vline & $12^\circ$.  &\vline & \\
$\Delta _{a}$ &\vline &  Full arm width measured normal to the arm.& \vline & 0.8 kpc  &\vline & \\
$v_{a}$&\vline &  Arm velocity with respect to the LSR. & \vline & $-50 \ {\rm km \, s}^{-1}$  &\vline & \\ 
\hline
    
    & & {\hspace{-2 cm}$^a$ Truncated power-law in \cite{ric16}.}&  & & &  \\
    & & {\hspace{-2 cm}$^b$ PDF for inner Galaxy clouds in \cite{miv17}.}&  & & &  \\
\hline\\
\end{tabular}
\end{table*}

\subsubsection{Formation and Distribution of GMCs}
\label{sec:sf}

The clustering of star formation events in space and time is driven by the fact that much of the
mass in GMCs, where stars form, is concentrated in the most massive clouds \citep{wil97,hey15}.  
We refer the reader to \citet{hen12} for a review of the physics of molecular clouds and of their formation.

For our standard case we adopt the GMC model of \cite{wil97} (hereafter WM97), with one important exception:
\citet{miv17} have shown that the surface density of the molecular gas in the Galaxy 
scales as $\exp(-R/H_R)$ with $H_R=2$~kpc, significantly less than the 3.5 kpc estimated by WM97. 
\citet{miv17} found that this distribution extends from 4 to 17 kpc. In order to approximately include the molecular gas
inside 4 kpc, we extend the exponential in to 3 kpc and assume that the density of molecular gas inside that is negligible.
The mass of molecular gas inside the solar circle, including He, is about $1.0\times 10^9$ M$_\odot$ ($\pm 30\%$)  according
to \citet{hey15}, just as estimated by WM97. 
In our model GMCs are created following a truncated power-law mass distribution:
\beq
\frac{d{\cal N}_c(M)}{d\ln M}={\cal N}_{\rm cu}\left(\frac{M_u}{M}\right)^{\gamma_c} ~~~~~~ (M_l\leq M \leq M_u),
\label{eq:WM97}
\eeq
where $d\caln_c/d\ln M$ represents the mass distribution of all the GMCs inside the solar circle. 
 In our standard case we assume the WM97 parameter values  $M_u=6 \times 10^6$ M$_\odot$, ${\cal N}_{\rm cu}=63$ and
$\gamma_c=0.6$, and assume that the actively star forming clouds have masses over $M_l=10^5\,M_\odot$ which
contain 80\% of the molecular mass.  The mass distribution of GMCs is somewhat uncertain,
and we discuss in Section 3.2.3 below the variation in the literature of the parameters $\gamma_c$, $\M_u$, ${\cal N}_{\rm cu}$,
test several other mass distributions, and determine the effect of these uncertainties on the determination of $\Gamma_{\app}$.  Because most of the
mass is in the largest clouds ($\gamma_c<1$), the results are insensitive to the value of $M_l$.

We represent the radial distribution of the GMCs as
\beq
\frac{d^2{\cal N}_c(M)}{dA\, d\ln M}=\frac{e^{(R_0-R)/H_R}}{A_\eff}\frac{d{\cal N}_c(M)}{d\ln M},
\eeq
where $R_0=8.3$~kpc is the radius of the solar circle and 
\beq
A_\eff=2\pi H_R^2 \left[\left(1+\frac{R_i}{H_R}\right)e^{(R_0-R_i)/H_R}-\left(1+\frac{R_0}{H_R}\right)\right],
\eeq
where $R_i$ is the inner radius of the distribution; this differs by a factor
$\exp(R_0/H_R)$ from the expression in \cite{mck97}. 
 For our adopted parameters ($H_R=2$~kpc, $R_i=3$~kpc,
and $R_0=8.3$~kpc), we have $A_\eff=760$~kpc$^2$. There are about 200 clouds more massive than $10^6\,M_\odot$
inside the solar circle,
containing about half the total mass of molecular gas; this corresponds to a surface density of 
$200/A_{\rm eff}=0.26$ GMCs kpc$^{-2}$.    
We assume that GMCs make stars for a period of 20 Myr \citep{bla91}, which we adopt as their lifetime.  The typical star in the mass range
$m=3-6$ has a lifetime of about 200 Myr (weighted by the PDMF) and therefore an observed age of about 100 Myr.
In that time, the total number of $M>10^6\,M_\odot$ GMCs that would be born is about $(100/20)\times 0.26$ kpc\ee~$=1.3$ kpc\ee. The mean distance between these associations is about 1 kpc. Since this is the distance that a star can travel
in 100 Myr at a typical velocity dispersion of 10 km s\e, we expect significant inhomogeneity in the surface distribution of these stars.
Recall from the discussion in Section \ref{sec:char} that a decrease in the number of $m=3-6$ stars relative to
the number of $m=1.5=3$ stars by 36\% could alter the inferred $\gap$ from 1.35 to 2. As a result, we expect significant
effects from the inclusion of clustering of star formation.

We simulate clouds with masses exceeding $10^5\,M_\odot$,
which contain 80\% of the mass; their surface density
is 1.5 kpc\ee. We assume that the formation rate of GMCs is proportional to the SFR; hence,
for a SFR that declines with time as $\exp(-t/\tsfr)$, the number of GMCs grows with look back time as $\exp(t'/\tsfr)$.
We adopt $t_{\rm SFR}=8.3\, \Gyr$, so that about 10000
are formed in the 37.5 kpc$^2$ simulation area  during 3 Gyr, but only 1800
of them have masses above $10^6$ \msun.
In the standard case the placement of the GMCs is random in the $x,y$ plane and at $z= 0$, but 
in Section \ref{sec:spiral} we consider the effect of spiral waves in concentrating the population of GMCs. 
 The scale height of GMCs is small 
($\sim 60$ pc--\citealp{hey15}) compared to the scale heights that the stars attain
after traveling through the ISM, so that the assumption of $z=0$ is justified.

\subsubsection{Star Formation in GMCs}

Stars are formed in the GMCs at a constant rate so that over the GMC lifetime a fraction $f_e$ of the mass $M$ of a particular
cloud
is converted  into stars.  
The average current rate of star formation in the Galaxy is $\simeq1.6\, M_\odot$/yr 
(\citealp{cho11}, \citealp{lee12}).  Assuming that the SFR scales with the surface density of GMCs,
we compute that  within the solar circle 
the star formation rate is 1.44 \msun yr$^{-1}$.  
Comparing this to the molecular mass $7.6\times 10^8$ \msun
in this region and utilizing the fact that GMCs lifetimes are 20 Myr, we find the average efficiency of star formation is $f_e \simeq 0.038$.
For simplicity we assume that the star formation efficiency is constant over the mass range $10^5$ to $6\times10^6$ \msun, and
we neglect star formation in lower mass clouds.
We have made simulations where for each GMC we allow for 1/2 standard deviation around the average star formation efficiency,
and find that the results for the $\Gamma_{\app}$ distribution have only minimal changes.  Although we form stars at random positions in the GMCs, the placement is not important since GMCs are 
very small compared to the interstellar volume
 over which we average.

The stellar  mass distribution in each star forming event is set by a fixed $\Gamma$ for $m>1.5$, and  we focus on stars of mass $m=1.5-6$. 
We have run simulations with $\Gamma= 1.2$, 1.35, 1.5 and 1.7.   We find (see Sec. 3.2.1 below) 
that if we plot the distribution of $\Delta
\Gamma_{\app} \equiv \Gamma_{\app} - \Gamma$, the results do not depend on $\Gamma$ in this range.    Therefore, we present
our results in terms of the distribution of $\Delta \Gamma_{\app}$, the deviation of $\Gamma_{\app}$ from the underlying $\Gamma$.

 {\subsubsection{Stellar Dispersal from GMCs into the ISM}
 
The effect of stellar clustering on the inferred IMF depends critically on the rate at which stars move from
their birth sites. We describe our method for determining the final location of each star in the Galactic plane
in Appendix A. We use data from \citet{nor04} and \citet{deh98} to fit the time-dependent expression
for stellar velocity dispersions given by \citet{aum16} for ages $\tau$ greater than 0.5 Gyr. For our standard
expression, we linearly interpolate from an initial dispersion of 6 km s\e\ \citep{aum16} to the values at 0.5 Gyr.
Each star is then given a velocity drawn at random from a Gaussian distribution with a dispersion equal to
the time-averaged dispersion for a star of that age, and its final position is determined.
To gauge the effects of uncertainties in the velocity dispersion, we also consider an upper limit on
the dispersion given by using the \citet{aum16} expression for $\tau < 0.5$~Gyr and a lower limit
given by setting the dispersion equal to its initial value, 6~km~s\e.
If the stars encounter
a boundary in the simulation during their lifetime, we adopt periodic boundary conditions to recover the stars that leave the simulation area
(see Appendix A).
At the end of the simulation (at 3 Gyr)  the stars are distributed in $z$ with the age dependent scale height given by Equation (\ref{eq:htSD03}).
 At this time we know the $x,y,z$ position and mass of each star in our simulation volume.  

\subsubsection{Determination of the Probability Distribution of $\Gamma_{\app}$}

 Knowing the position and mass of each star, we then place small (compared with the simulation size) spherical volumes of 
 radius $r_\obs$, based on observational limits, 
 within each simulation, and run numerous simulations to
add to the statistics.  Inside the volumes of radius $r_{\obs}$ we 
determine $\calr$ by counting the stars in the mass range $m=1.5-3$ and rationing that number with
 the number of stars with $m=3-6$.   
Since we have included the SFH and vertical diffusion in
the simulation, this value of $\calr$ is $\calr_{12}$, and we can use the appropriate curve in Figure 1 to infer $\Gamma_{\app}$.
This value of $\Gamma_\app$ differs from the underlying $\Gamma$ only because of inhomogeneities in the 
relative 
densities of low- and high-mass stars.   
These inhomogeneities are mainly caused by the fact that the shorter-lived, high mass stars do not move
rapidly enough from their localized birthsites (GMCs) in their lifetimes to uniformly populate the
Galactic plane.  Each  volume then can have an excess or a deficit of massive stars relative to low
mass stars, and there is therefore a range of values of $\gap$ even though $\Gamma$ is fixed.
In what follows, we examine the dependence of the distribution of values of $\Delta\gap=\gap-\Gamma$
on the observational volume, the underlying $\Gamma$, the dispersion speed of the stars, and the initial 
spatial concentration of the stars.  We also examine the effect of spiral density waves on the distribution of the GMCs, and the subsequent distribution of $\Delta \Gamma_\app$.

\subsubsection{Uncertainties in the Simulations}

We emphasize that our simulations should be considered as examples, and not a precise representation
of conditions at the solar circle.   We have chosen plausible parameters and have endeavored to obtain a good
estimate of the inhomogeneous distribution of stars in the plane of the Galaxy, and of the variations
in the SFR per unit volume as a function of position along the solar circle.  However, there are many uncertainties:
The mass distribution of GMCs,
their lifetimes, and the star formation efficiencies inside GMCs with different masses are all uncertain.
The velocity distribution of newly formed stars as a function of time
 is uncertain, as they
proceed from the random velocities incurred during their formation in GMCs to the velocities
they achieve after many gravitational encounters with molecular clouds. In our initial model
we assume that GMCs form at random positions in the plane.  However, 
 in Section 4 we describe the possible effects caused by spiral arms. This treatment is also
 exemplary, because
the number and strength (i.e., the ratio
of the SFR per unit area in the arms versus that in the interarm) of these arms is uncertain,
as is the exact position of the Sun with respect to these arms. Probably the two most significant uncertainties are the mass distribution of GMCs 
and the effect of spiral arms in affecting the spatial distribution of GMCs.   We test the effect of these uncertainties
in Sections 3.2.3 and 4 below.

\subsection{Results}
\label{subsec:results}

In all the results presented below, we carried out 360 simulations in which we placed 25 independent observing spheres
of radius $r_\obs$  (our standard value is $r_{\obs}=200$~pc but we present cases where the radius is varied) centered 
in the mid-plane and separated by 1 kpc,
providing 9000 different independent values of $\Delta\gap$. The standard of 200 pc is  chosen because both
Dawson \& Sch\"oder (2010) and Rybizki \& Just (2015) use observations samples with $r_{\obs}=100- 200$~pc.
However, Czekaj et al (2014) use a sample with apparent magnitude up to V=11, which corresponds to a distance of
about 300-400 pc for the lower mass stars in our sample.  Therefore, we include analysis up to $r_{\obs}=600$~pc
in this section and even larger in the next section.   Note that our simulation uses volume limited samples, not
magnitude limited samples.

All the results presented in this paper are for observational spherical volumes centered at the mid-plane. Off plane results require a reevaluation of $F(t^{'} , r_{\obs})$ to infer 
$\Gamma_{\app}$ from the value of $R$ within an off-plane sphere (i.e. a new Figure 1). However, if the sphere is placed at a distance above the plane that is small compared to  $r_{\obs}$ and to the 
scale height of high-mass stars, one expects negligible differences compared to the results obtained with the spheres centered in the mid-plane. 
The Sun is located about 15-25 pc above the mid-plane (\citealp{dri01}), a 
distance that is small compared to  $r_{\obs}$ and to the scale height of high-mass stars, both larger than 100 pc.  Therefore we ignore this effect in the simulations that follow.

In this section, we form GMCs at random positions around the solar circle.   In Section 4 we treat GMCs forming preferentially in spiral arms.

\subsubsection{Probability of finding $\Delta \Gamma_{\app}\ge +.35$ for $r_\obs=200$ pc, independent of $\Gamma$}

Figure 2 shows the probability distribution 
of the inferred $\Delta \Gamma_\app$ for four different values of $\Gamma$.
For the case under consideration ($r_\obs=200$~pc), the mean and median value of $\Delta \Gamma_{\app}$ are about
+0.1 and the dispersion is about $\sigma\simeq 0.45$.
In other words, the inhomogeneous distribution of stars resulting from clustered
star formation as described in Section \ref{sec:sf} causes the inferred $\Gamma_{\app}$ 
to be typically about
0.1
higher than the underlying $\Gamma$.
This conclusion should depend only weakly on the parameters we have adopted
for the SFH ($\tsfr=8.3$~Gyr) and
stellar scale heights given in Equation (\ref{eq:htSD03}) since they are accounted for in our relation
between the ``observed" $\calr$ and the inferred $\gap$--there would be no dependence in the absence of
clustered star formation, since then $\gap=\Gamma$, ignoring the small statistical fluctuations due
to finite sample size.

In addition, we can estimate
the likelihood that $\Delta \Gamma_{\app}$ is greater than any fixed value.   
For example, the probability that
$\Delta \Gamma_{\app} > 0.35$ -- i.e., probability that the observer/modeler obtains $\Gamma_{f,\app} > 1.7$, 
when in fact the
underlying $\Gamma= 1.35$ -- is $\sim 30\%$.     
 We shall discuss the effect of
spiral arms below, but in the absence of their effect, 
these results suggest that the observation of $\Gamma_\app \simeq
1.7-2.1$ favors a somewhat higher value of $\Gamma$ than 1.35, although
there still is a significant possibility that $\Gamma=1.35$. 

Finally, we find that the distribution of values of $\gap$
is independent of the underlying slope, $\Gamma$.  
In the rest of this section, we therefore present results as functions of $\Delta \Gamma_{\app}$.

\begin{figure}
\hbox{\includegraphics[width=\columnwidth]{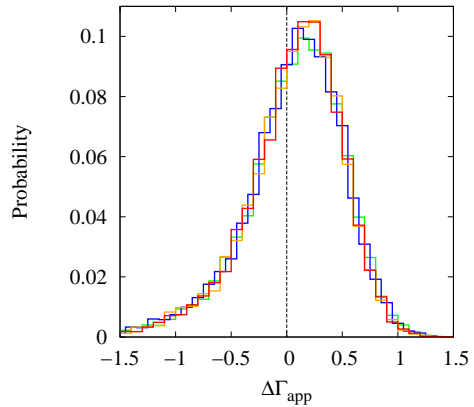}}
\caption{
Probability distribution of $\Delta \Gamma_\app$ for four different values of the underlying $\Gamma$.  Green is $\Gamma=1.2$, blue is $\Gamma=1.35$,  orange is $\Gamma= 1.5$, and red is $\Gamma=1.7$. 
The bin size is $\Delta\gap=0.05$.
The plots are essentially the same, showing that  
probability distribution of $\Delta \Gamma_\app$ does not depend on $\Gamma$. We assume here the standard stellar random velocity distribution and $r_\obs=200$ pc. 
}
\label{fig-2} 
\end{figure}

\subsubsection{Dependence of $\Delta \Gamma_{\app}$ on the Stellar Velocity Dispersion}

Before proceeding further, it is important to show that our results are not very sensitive to the
velocity dispersion, as long as the values remain in reasonable bounds.   Figure 3 presents
three representations of the time dependent velocity dispersions of the stars
for the case where $r_{\obs}= 200 $ pc (see Appendix A for details of our treatment of
velocity dispersion). Stars are
accelerated from their initial random speeds by gravitational forces in
the Galactic plane (e.g., passage of spiral arms or encounters with GMCs). Therefore, the minimum velocity
dispersion of a star would be to hold the dispersion constant in time with the initial
dispersion, which we have taken to be 6 km s$^{-1}$ from \citet{aum16}. 
We take the time-dependent
velocity dispersion given by \citet{aum16}, $\sigma_{\ABS}(\tau)$, to be the maximum since
it is observed to be too large at early times--the radial velocity at $\tau=0$ is more than twice
the observed value (see Appendix A).   
The standard 
velocity distribution is intermediate between the minimum and maximum dispersions, with
the velocity dispersion increasing linearly with $\tau$
from the lower to the upper limit during the first 500 Myr and matching $\sigma_{\ABS}(\tau)$ thereafter.
As expected, the minimum velocity dispersion has a slightly 
broader probability distribution as a function of $\Delta \Gamma_\app$,
since the slower stars maintain a larger spatial inhomogeneity.  
However, the difference between the three cases is small
because the random stellar speeds ($\sim 10$ km s$^{-1}$) in all three cases are quite similar.  In addition,
recall that GMCs are forming at random positions constantly to replace GMC that have died.   This places
stars more uniformly in the Galactic plane than if the sites of star formation were frozen in time.
Figure 3 shows that our results are not sensitive to the standard velocity dispersion we have chosen.
The overall conclusion is that for reasonable values of the velocity distribution of the stars, migration is not enough to eliminate the effect of 
the clumpy nature of the star formation process.

\begin{figure}
\hbox{\includegraphics[width=\columnwidth]{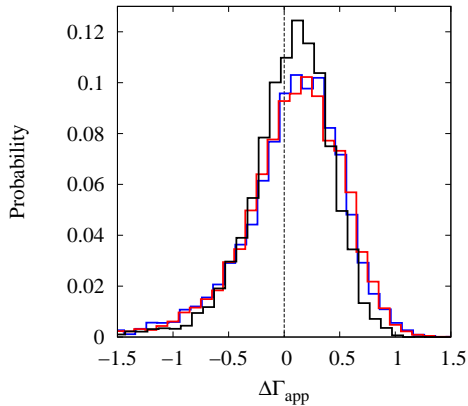}}
\caption{ Distribution of $\Delta \Gamma_\app$ for our three velocity dispersions of stars, assuming $r_\obs=200$ pc.
   Red is the minimum dispersion of
6 km s$^{-1}$ in U and V.
Black is the maximum dispersion as given in Aumer et al. (2016;
see text).  Blue plots the probability distribution for the standard case, in which 
the velocity dispersion linearly increases from 6 km s\e\ to the \citet{aum16} result
during the first 500 Myr.
}
\label{fig-3} 
\end{figure}

\subsubsection{Dependence of $\Delta \Gamma_{\app}$ on the Concentration of GMCs}

\begin{figure}
\hbox{\includegraphics[width=\columnwidth]{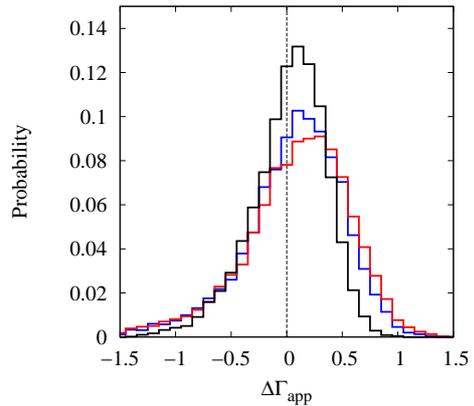}}
\caption{
Probability distribution of the deviation $\Delta \Gamma_{\app}$ from the underlying IMF slope  for three  different concentrations  
of the star formation events, corresponding to three different mass distributions of GMCs.
The blue histogram corresponds to our standard case, where the mass distribution is from WM97.
The red curve corresponds to the mass distribution of Rice et al (2016, see text).   This distribution has more molecular mass in
fewer, larger GMCs and therefore is the most concentrated and results in the broadest distribution in $\Delta \Gamma_{\app}$ as expected.
On the other hand, the \citet{miv17}  distribution puts mass into a greater number of clouds, leading to a narrower distribution in
$\Delta \Gamma_{\app}$ (black curve).
Other parameters are $r_{\obs}=200$ pc and standard velocity distribution. 
}
\label{fig-4} 
\end{figure}

We next examine the dependence on the concentration of star formation events.  
 The distribution by mass of GMCs is uncertain, and there have been several studies in the recent literature with
somewhat different results. Clouds outside the solar circle are smaller than those
at or inside the solar circle (\citealp{hey15}), so we focus on the latter.
The data used by WM97 (e.g., \citealp{sol87}) assigned only 40\% of the molecular gas to clouds, and WM97 then
assumed that the remaining 60\% was in clouds of the same mass distribution, but were too cold to appear
in the catalogs. We note as justification that locally all the molecular gas is in clouds. Subsequently,
\cite{ros05} found $\gamma_c =0.53$, $N_{\rm cu}/\gamma_c=27-36$, and $M_u=(3-4)\times 10^6\, M_\odot$
based on the data of \citealp{sol87}, whereas
\cite{ric16} found $\gamma_c =0.6\pm 0.1$, $N_{\rm cu}/\gamma_c=11\pm6$, and $M_u=(1.0\pm0.2)\times 10^7\, M_\odot$.
These parameters were introduced and discussed in Section 3.1.1 above.
(Note that these workers assign only a fraction of the molecular mass to clouds, whereas
WM97 assumed it was all in clouds.) On the other hand,
\citet{miv17} did not find a cutoff distribution, but instead found $\gamma_c=2$ at very high masses;
furthermore, they assigned 98\% of the CO emission to clouds.
One thing 
all
these 
distributions have in common is that most of the molecular mass in the distribution
is concentrated in the most massive clouds, which are not abundant and are therefore widely spaced.   Hence, if all
the molecular mass is assigned to these cloud distributions, then the results will not differ significantly from our standard model
using WM97.  We show in Figure 4 the effect of these different
GMC distributions on $\Delta \Gamma_{\app}$ when all the molecular mass is assigned to these distributions.  
 If the 
 $\sim$ 60\% of the CO emission not assigned to clouds by \citet{sol87}, for example,  
 is in more diffuse clouds that are not actively forming stars,
then there would be even fewer star-forming clouds than in our standard WM97 model, and 
the more widely dispersed sites of star formation would
produce greater deviations in $\Gamma_{\app}$ from $\Gamma$.

Figure 4 shows the expected broader distribution
in $\Delta \Gamma_\app$ when stars form in more concentrated spatial locations
(\citealp{ric16} case) compared to the \citet{miv17} case which distributes the mass in a greater number of smaller clouds.    
More localized birth sites lead to a final
stellar spatial distribution that is much more inhomogeneous (especially for the shorter-lived high-mass stars), and thus 
the inferred $\Delta \Gamma_\app$  shows a broader distribution and, as noted 
above, a shift ($\sim +0.1$) toward positive values
(that is, $\Gamma_{\app}$ greater than the underlying $\Gamma$). Although not
evident in Figure 4, the narrowing of the distribution in the \citet{miv17} case has a significant effect on
the probability that $\Delta \Gamma_{\app} > 0.35$, or, in other words, that $\Gamma_{\app}>1.7$ when 
the underlying $\Gamma= 1.35$.    Assuming $\Gamma= 1.35$, the WM97 case, the \cite{ric16} case
and the \citet{miv17} case have the following probabilities that $\Gamma_{\app}>1.7$: 0.27, 0.31, and 0.15.
We therefore conclude that the uncertain distribution of GMCs by mass leads to a $\sim 15-30$\% chance that
$\Delta \Gamma_{\app} > 0.35$ if GMCs are formed randomly in the plane of the solar circle.

\subsubsection{Dependence of $\Delta \Gamma_{\app}$ on Observing Volume}

\begin{figure}
\hbox{\includegraphics[width=\columnwidth]{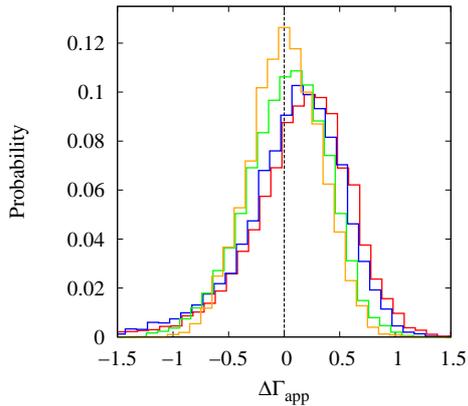}}
\caption{
Probability distribution of $\Delta \Gamma_\app$ as a function of observing volume.
The red, blue, green and orange curves correspond to the cases when the ratio $\calr $ is measured within a sphere of radius 100, 200, 400 and 600 pc respectively.
As expected, as  $r_\obs$ increases the distribution narrows around  $\Delta \Gamma_{\app}= 0$, i.e., $\Gamma_{\app}$
approaches $\Gamma$.
}

\label{fig-5} 
\end{figure}

Figure 5 shows the dependence of the probability distribution of $\Delta \Gamma_\app$ on the observing volume for the standard
case with the standard velocity
distribution.
Recall that our approach, using Figure 1 to determine $\Delta \Gamma_\app$ from the ratio $\calr $, has corrected for the vertical differences in massive and low mass stars as well as the SFH.   Therefore, the 
 narrowing of the distribution 
 and the shift to an average value $\Delta\gap\simeq 0$
 as $r_\obs$ increases is solely due to the larger volumes averaging out  
 the $x,y$ spatial inhomogeneities  of the more massive stars. 
 The mean and median values of $\Delta \Gamma_{\app}$ for the cases corresponding to $r_\obs=$100, 200, 400 and 600 pc are respectively 
 (0.13, 0.17), (0.09, 0.13), (0.02, 0.04) and (0.02, 0.02).   
 This demonstrates that the effect of inhomogeneity systematically  increases $\Gamma_{\app}$ above the underlying $\Gamma$ 
 as the observing volume shrinks.
The percentage of simulations with $\Delta \Gamma_\app$ in the range -0.15 to +0.15 are respectively 24\%, 26\%, 30\% and 34\%.  
 The percentages of simulations with $\Delta \Gamma_\app>0.35$  (so that, for example, measures of 
 $\Gamma_{f,\rm app}>1.7$
 actually correspond to an underlying $\Gamma= 1.35$)  are 33\%, 30\%, 20\%, and 17\%.  Thus, for example, with
 $r_{\obs}= 200 $ pc it is more likely to infer $\Gamma_{\app}> \Gamma+0.35$ than it is to infer that 
 $\Gamma_{\app}$ lies within $\pm 0.15$ of $\Gamma$.
In another interesting example of these results, for a volume sample of $r_\obs=$200 pc with random placement of the Sun
along the solar circle (ignoring the effects of spiral density waves), the chances of 
inferring $\Delta \Gamma_\app$ within $\pm 0.25$ of the underlying $\Gamma$ (42\%) is similar to   
the probability of inferring $\Delta \Gamma_\app>0.25$ (43\%).

 The most important features in Figures 2-5 are: (i) the broad dispersion in  $\Delta \Gamma_\app$ which is the
main point of this paper, (ii) the shift in  $\Delta \Gamma_\app$ toward positive values
(i.e., $\Gamma_{\app}$ larger and thus a steeper apparent IMF), and (iii) an asymmetry between negative and positive
values with a long tail on the negative side (cases with a flatter IMF and more massive stars).  The shift to positive
values of  $\Delta \Gamma_\app$ is due to the short lifetimes of massive stars and the low space density of young
star forming regions.   Most of space has a deficit of massive stars compared to the density these stars would have
if star formation were homogeneous in space.   Therefore, for a fixed observing volume, there is greater chance of
being in a region of deficit than of excess of massive stars.  Similarly, the tail on the negative side is created by
those few volumes that fall around active star formation regions.   Both the shift and the asymmetry decrease as the
observing volume increases, because larger volumes encompass both regions of deficit and excess.

The standard deviations of the $\Delta \Gamma_{\app}$ distributions decrease only moderately as $r_{\obs}$ increases from 100 to 600 pc (from
0.45  to 0.32).\footnote{
The standard deviations of the $\Delta \Gamma_{\app}$ distribution continue to decrease as the sampled region increases, 
reaching the value of 0.05 for the whole simulation area.
}
The modest amount of narrowing with observing radius  might seem surprising, but recall that most stars form in the largest GMCs that
 are $\sim 1$ kpc apart, and that a significant fraction have slow diffusion speeds so that significant variations can still occur on
 the 600 pc scale.    In addition,  the distribution depends on the local star 
 formation history in the general location of the observing volume.   Although the {\it average} SFH at the solar circle
  declines somewhat in our standard case over the last 3 Gyr, the localized nature of star formation
 means that
 there are significant fluctuations in the SFR over smaller volumes and over timescales of order the lifetime ($\sim 100$ Myr) of the massive $m=3-6$ stars.  We study these temporal fluctuations next.

 \subsubsection{Fluctuations in the SFR Over Small Volumes}

\begin{figure}
\hbox{\includegraphics[width=\columnwidth]{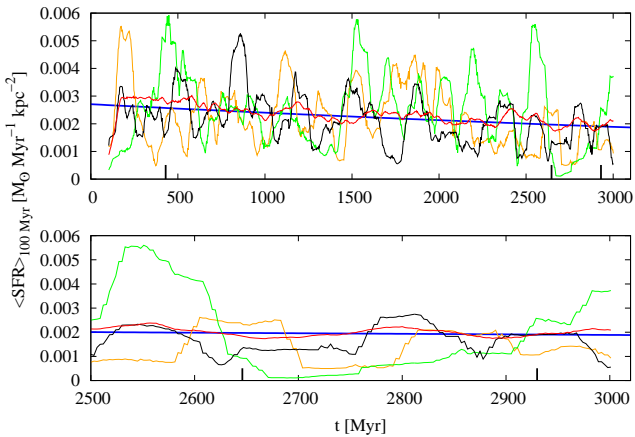}}
\caption{ 
Evolution of the star formation rate averaged over the previous 100 Myr, $<$SFR$>_{100 \rm Myr}$, for a particular simulation.
As in all simulations, stars form in widely spaced GMCs.
The current epoch is $t\equiv t_0=3000$ Myr. The lower panel is an expanded version of the upper panel for the last
500 Myr. The vertical ticks on the $x$-axis indicate the survival times (MS stellar lifetimes) into
the past for stars of 1.5, 3 and 6 $M_\odot$.
The red curve corresponds to the $<$SFR$>_{100 \rm Myr}$ occurring in the whole simulation area ($0<x<25$ kpc and $0<y<1.2$ kpc).
The green, black and orange curves correspond to the $<$SFR$>_{100 \rm Myr}$ in three 1.2 kpc$^2$  regions of the simulation area,
respectively in the x-axis sections 4-5 kpc, 9-10 kpc and 16-17 kpc.
The blue curve corresponds to the smooth average evolution of the SFR at the solar circle, or SFR= SFR$(t_0) \, e^{t^\prime /t_{\rm SFR}}$ 
where $t^\prime \equiv t_0-t$ is the look back time 
in Gyr.
The simulation parameters are  $t_{\rm SFR}=8.33\,\Gyr$ and standard velocity dispersion.
}
\label{fig-6} 
\end{figure}
 
 \cite{elm06} have discussed the effect of peaks and lulls in the SFR 
 on the inference
 of the high mass IMF from the PDMF.   For example, if a region has suffered a lull over the past 100 Myr, there will be a deficit of
 short-lived, massive stars and the inferred slope of the high mass IMF will be steeper.
Figure 6 displays the fluctuations in the star formation rate  $<$SFR$>_{100 \rm Myr}$ averaged over the previous 100 Myr.  If the SFR
is averaged over the entire 30 kpc$^2$ area of the simulation
(red curve), we see that the fluctuations are not large.  However, when the
SFR is averaged over smaller 1.2 kpc$^2$ areas, the SFR changes significantly over $\sim 100$ Myr timescales. 
Note that the black and orange curves fall significantly below the
global average SFR whereas the green curve falls significantly above the global average.   The underproduction
of $m=3$ stars in the black and orange cases should lead to a higher value of $\Gamma_{\app}$ whereas the
overproduction seen in the green case should lead to a low value of $\Gamma_{\app}$ compared to the underlying $\Gamma$.
In fact, in this particular simulation where we have chosen an underlying $\Gamma=1.35$ the values of the ratio 
$\calr $ within 600 pc from the points at x=4.5 (green), 9.5 (black) and 16.5 (orange) kpc 
at $t_0$ are respectively 11.7 ($\Gamma_\app$=0.96), 21.6 ($\Gamma_\app$=1.83) and 21.5 ($\Gamma_\app$=1.82),
in accordance with the above prediction.

\section{Effect of Spiral Arms on the PDMF}
\label{sec:spiral}

If the Sun were located in a random position at the solar circle, we would then conclude from the above
simulations that observations with a limited observing volume
would have a roughly 15-35\% chance of finding $\Gamma_\app \simeq 1.7-2.1$ if the underlying $\Gamma= 1.35$.
 However, the Sun is not at a random position: it currently resides in
an interarm region
where the star formation rate is low, and as a result the PDMF is altered in such a way as
to suggest a steeper IMF (\citealp{elm06}; in our terminology, this leads to a positive $\Delta\gap$).
As first noted by \cite{dri01}, observations of near- and mid-IR stellar tracers indicate the presence of only two
spiral arms. However, observations of the gas and dust in the radio and far-IR indicate four arms
(e.g., \citealp{geo76,koo17}). We therefore follow \citet{ben08} and \citet{rob12} and assume that there are two strong arms
and two weak ones. The Sun has left the weak Sagittarius-Carina arm and will encounter the strong Perseus arm.
In our spiral arm model we include star formation that occurs in the interarm region but with
a lower intensity than in the strong and weak arms.   We neglect the Orion Spur, a minor arm that crosses
the region where the Sun is located (\citealp{dri03}), and consider it as part of the low intensity star formation that
occurs in the interarm region of our model.
As discussed in Appendix B, the speed of the arms with respect to the Local Standard of Rest (LSR) is about
-50 km s\e; with 4 arms, this implies that the solar neighborhood 
is overtaken by an arm every $\sim 260$ Myr. 
However, due to the finite width of the arms and the small pitch angle (see Appendix B) the time spent in the interarm region is reduced to $\sim 180$ Myr.
The star formation rate per unit area is larger 
in the arms than in the interarm regions, so the Sun has experienced a relatively low star formation rate for 
$\sim 90$ Myr
preceded by a higher rate when the Sun lay in an arm.   
The lower mass stars ($m\sim 1.5$) live long enough to diffuse
through the solar circle and therefore have a relatively uniform population when comparing arm and interarm.  However, the
higher mass stars ($m>3$) have shorter lives and are more concentrated in the arms, which leads to larger values of
$\Gamma_\app$ in the interarm region compared to the underlying $\Gamma$.

To consider the effect of the spiral arms, the only change we make is to assume that the probability of formation of GMCs is enhanced in the arm regions.
We fix this enhancement by the fraction, $F_S$, of the  GMCs that are formed in the two strong arm regions and the fraction, $F_W$, of GMCs formed in the two weak arm regions. 
The properties of the spiral arm model that we adopt are described in Appendix B: the number of arms, 4,  
the star formation fractions, $F_S=0.5$ and $F_W=0.2$, the pitch angle $\alpha=12^\circ$, 
the arm width normal to the arm, $\Delta _{a}=0.8$~kpc, and the arm velocity with respect to the 
LSR, $v_{a}=-50$~km s\e, as noted above. 
Note that the thickness of the arm along a circular orbit is
$\Delta_x=\Delta_a/\sin\alpha=3.8$~kpc.  The key parameter in considering the effect of spiral arms is $F_I\equiv 1-F_S-F_W$, the fraction
of GMC formation in the interarms.   As discussed in Appendix B, observations of external galaxies suggest that the
range of $F_I$ is 0.1-0.6.    For our standard case mentioned above, we take $F_I=0.3$, but we also treat a case with
$F_I=0.5$.    Smaller values of $F_I$ lead to more suppression of the SFR in the interarm over the past $\sim 90$ Myr, and thus
a larger value of $\Gamma_{\app}$.    

\begin{figure}
\hbox{\includegraphics[width=\columnwidth]{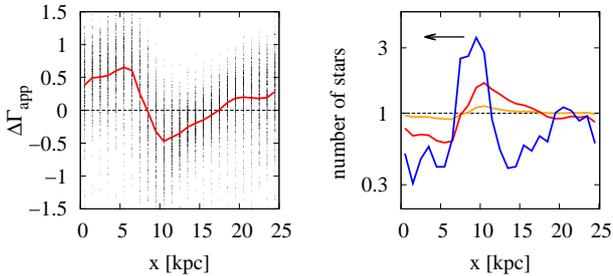}}
\caption{Effect of the spiral wave density for 2 strong and 2 weak arms (see text). 
Left panel: the dots represent all the $\Delta \Gamma_\app$ values in 360 simulations, taking $r_{\obs}= 200$ pc
as the radius of our observing sphere. 
In each simulation 25 spheres are placed in the simulation area at intervals of $\Delta x= 1$ kpc along
our 25 kpc simulation length.
The red line connects the mean of the values of $\Delta \Gamma_\app$ in the 360 simulations at each $x$ point.
The right panel shows the average number of stars that are within a sphere at position $x$ in the 360 simulations normalized to the average number of stars per sphere. 
The blue, red and orange curves correspond respectively to stars of masses $m>10$, $3\leq m < 6$, and $1.5 \leq m < 3$.
The arrow on the strong arm denotes the direction of the spiral arm with respect to the stars.
The fraction of GMC formation in the interarm is $F_I=0.3$; see text for other arm parameters.
}
\label{fig-7} 
\end{figure}

Figure 7 shows the effect of the modulation of the SFR due to the spiral wave passage for the standard spiral arm case
described above ($F_I=0.3$).
These parameters imply that the relation of the star formation rate per unit area in the strong-arm, weak-arm and interarm
regions  is respectively 8:3:1.
The results shown correspond to 360 simulations done for the same set of parameters. 
Each simulation differs in the random placement (size, time and position) of the star formation events, as well as 
the in the $x,\,y$ velocities of the stars at birth and their $z$-positions at the end of the simulation.
As in all our simulations, each simulation provides the star counts at the final time (at $t=t_0=3$ Gyr) in 25 observational spheres centered at 
$z=0$, $y=y_{T}/2$ and separated from each other by 1 kpc in the $x$ direction.
The front-side (inner edge) position 
of the strong arm at $y=y_T/2$ at the final moment of the simulation is $x_{fs}(t_0,y_T/2)=6.5$ kpc,
whereas the front side position of the weak arm is located at $x_{fw}(t_0,y_T/2)=6.5+x_T/2 = 19$ kpc.  The arms move to the left
in Figure 7 with respect to the stars orbiting the Galactic Center.   Because our simulation area covers only 25 kpc (or 
about halfway) around
the solar circle, only two of the four arms are within Figure 7. 

The observational spheres located at the left of the front-side of the strong arm are those that have spent the most time
without being crossed by
a strong arm, and therefore are the ones where the  population of the 3 to 6 M$_\odot$ stars is most  depleted.
This is evidenced in the right panel of Figure 7, where the normalized abundance of stars in three stellar mass ranges ($m>10$, $3<m<6$ and $1.5<m<3$) is shown.
The stars with masses $m>10$ display a normalized abundance (the blue curve) that closely follows the arms since their lifetimes are shorter than
$\Delta_x/v_a$. This creates the ``beads on a string" appearance of external spiral galaxies.
The stars in the mass range $3<m<6$ have lifetimes between 357 Myr and 64 Myr, whereas the time between successive passages of the 
spiral arms is $\sim 260$ Myr, so one expects a  depletion of these stars between the arms. 
As shown in Figure 7 the amplitude of 
the spatial fluctuation of the abundance of these stars (the red curve) is considerably smaller than for the massive  $m > 10$ stars but larger 
than the less massive stars in the range $1.5<m<3$, whose lifetimes span from 2570 Myr to 354 Myr. 
The amplitude of the spatial fluctuation decreases as the mass of the considered stars decreases because the corresponding increase in average age implies that lower mass stars
(i) sample further back in time and therefore include stars formed in 
previous arm passages; and
(ii) spread out farther from their birth sites due to their random velocities.
As expected, to the left of 
the strong arm
the population of the 3 to 6 $M_\odot$ stars is more depleted than the  population of the 1.5 to 3 $M_\odot$ stars,
and consequently in this region $\Gamma_\app$ is steeper than the underlying $\Gamma$. 
Note also that GMCs form stars during a period of $\tau_{\rm GMC} \approx 20$ Myr and therefore the star formation is still enhanced along a distance 
$v_a \times \tau_{\rm GMC}\sim 1$ kpc behind the back-side of the arm.

The left panel of Figure 7 shows the 360 values of $\Delta \Gamma_\app$ in each sphere as well as the their local average (the red curve).
It is clear that the imposed variation of the SFR has an important effect 
on the local average of $\Delta \Gamma_\app$, but it is also clear that the dispersion of the $\Delta \Gamma_\app$ values around the local 
average (due to the clustered star formation) is even larger than the departure of the local average from zero  (that is, the
departure of $\Gamma_{\app}$ from the underlying slope $\Gamma$).

\begin{figure}
\hbox{\includegraphics[width=\columnwidth]{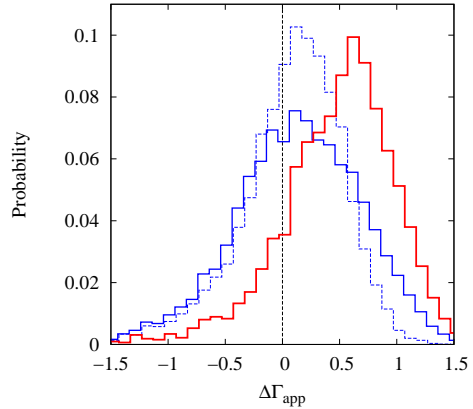}}
\caption{
Distribution of $\Delta \Gamma_\app$ when the effect of spiral arms in concentrating the GMCs is considered.
In this model there are four arms, and 50\% of the SFR occurs in the two strong arms, 20\% occurs in the two weak arms,
and 30\% occurs in the interarm regions.
The blue histogram includes all 9000 values of $\Delta \Gamma_\app$ for all (arm and interarm) locations.
The red histogram includes only the $\Delta \Gamma_\app$ values in the observational spheres located in the interarm 
range $-2<x<7$ kpc in Figure 7 
(see text).
The dashed blue histogram corresponds to the standard case without the effect of spiral arms, as shown in Figures 2-5. 
Simulation parameters are 
$r_\obs=200$ pc, standard velocity distribution, and arm parameters given in text.
}
\label{fig-8} 
\end{figure}

The blue curve in Figure 8 displays the overall (arm and interarm) probability distribution of $\Delta \Gamma_\app$ values for our 
standard 4-arm model with an observing radius of $r_\obs= 200$ pc and our standard stellar velocity distribution.
Comparison of the overall $\Delta \Gamma_\app$ distribution with the case without the GMC concentration in the spiral arms (dashed blue curve) indicates that 
the spiral wave has a significant effect on the probability distribution, especially enhancing the probability in the high tail due to the depletion 
of high mass stars in the interarm regions. The standard deviation increases from 0.45 to 0.55 when GMC concentration in the spiral arms is considered.
 The red histogram represents the distribution of the $\Delta \Gamma_\app$ values only 
in the nine  spheres in the region that precedes the strong arm. 
Note that due to the periodic nature of our four arms model the spheres located in the range $-2<x<7$ kpc correspond
to the spheres located in the ranges $23<x<25$ kpc and $0<x<7$ kpc in figure 7.
At the final time of the simulation these spheres have left the the weak arm from 0 to 
180 Myr ago  and are 
being approached by the strong arm.
As shown in Figure 7, in these spheres the average value of $\Delta \Gamma_\app$ is above the average for the
entire solar circle. In fact, in this interarm region the peak in the distribution occurs at $\Delta \Gamma_{\app} \sim 0.65$,
corresponding to $\gap=2.0$ when the underlying $\Gamma= 1.35$.

\begin{figure}
\hbox{\includegraphics[width=\columnwidth]{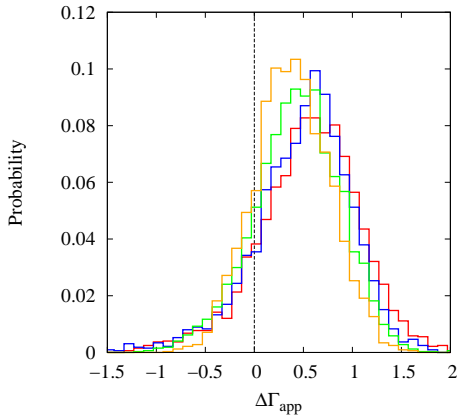}}
\caption{ The same as Figure 8 but for different sizes of the observational volume.  The red, blue, green
and orange curves correspond to $r_{\obs}= 100$, 200, 400 and 750 pc, respectively.  For clarity only the  histograms for $\Delta 
\Gamma_{\app}$  in the  interarm region are shown.
}
\label{fig-9} 
\end{figure}

 Since the Gaia mission has the ability to sample volumes in the interarm region at the solar circle, Figure
9 shows the dependence of the spiral arm model to the volume sampled, if the volume is kept in the interarm region.
Because of the large distances between the arms, the effect of the spiral arm does not depend strongly on $r_{\obs}$
 provided that $r_\obs$ is small compared to the distance to the corotation radius (we have assumed that $v_a$ is constant
in these simulations.
The mean and median values of $\Delta \Gamma_{\app}$ for a given volume are the same to within 2\%
so we just give mean values here.   For $r_{\obs}= 100$, 200, 400, and 750 pc we find: mean $\Delta \Gamma_{\app}= $+0.52,
+0.47, +0.41 and +0.36; standard deviations are 0.55, 0.49, 0.44, and 0.38; probabilities that $\Delta \Gamma_{\app} > 0.35$ are
0.67, 0.64, 0.57 and 0.51; probabilities that  $\Delta \Gamma_{\app} > 0.65$ are 0.42, 0.38, 0.30, 0.23.   In short, for observing
radii $r_{\obs}= 100$ to 400 pc, there is a 67\% to 57\% chance to obtain $\Gamma_{\app} > 1.7$ if the underlying $\Gamma= 1.35$
for our standard model of spiral arms at the solar circle.  We look at results for other models below.

The purpose of this paper is to examine the observer/modeler determination of $\Gamma_{f,\app} > 1.7$ in studies
of local field PDMFs and how these values might be larger than the underlying $\Gamma$.  Therefore,  we present in Figure 10
the results of our simulations in a different way.    We plot the probability of determining $\Gamma_{\app}>1.7$
as a function of the underlying $\Gamma$ for five samples of local PDMFs on the solar circle. 
(1) The average of all positions along the solar circle, including all arm and interarm positions.
(2) Regions in the strong arm (spheres located in the range $7 < x < 11$ kpc in Figure 7).
Here, the enhanced population of GMCs and of star formation leads to lower probabilities that $\Gamma_{\app}$ will
be higher than the underlying $\Gamma$, since these regions are well populated with massive stars.
(3) The interarm region (spheres located in the range $-2 < x < 7$ kpc in Figure 7) that precedes the strong arm.
 Here, although low-mass stars have had time
to migrate from arm locations, high-mass stars have had less time due to their shorter lifetimes and so
they are underrepresented, leading to higher $\Gamma_{\app}$ values in these interarm regions.
(4) The sphere centered at $x=5.5$ kpc where the average value of $\Gamma_\app$ reaches its maximum.   
This curve provides an upper limit for this particular model of spiral arms at the solar circle.
At the final time of the simulation this sphere has left the weak arm about 180
Myr ago and is being approached by the strong arm.
(5) The sphere centered at $x=10.5$ kpc where the average value of $\Gamma_\app$ reaches its minimum.
At the final time of the simulation this sphere is just leaving the strong arm.
This curve provides a lower limit for this particular model of spiral arms at the solar circle.
Finally, (6) The Sun is in an interarm region between the Sagittarius-Carina and the Perseus arms (e.g., \cite{val05}),
and we assume that the Sun is located at half way between the weak and the strong arms in the sphere centered at $x=1.5$ kpc. 
Comparing the 
orange curve (for the assumed position of the Sun, which is close to the brown curve for the average over the interarm region)  
to the black (average over solar circle) curve, we conclude that
the interarm location of the Sun can significantly (by a factor $\sim 1.5 -2$) raise the likelihood that a 
local measurement of $\Gamma_\app$ will exceed 1.7 when the underlying universal slope $\Gamma$ is
less than this value. 
The median value of $\Delta \Gamma_{\app}$ for the $-2 < x < 7$ kpc interarm region is 0.51. As  result,  for example,
if  the underlying $\Gamma= 1.35$, there is approximately 50\% chance that a local measurement
with observing radius 200 pc will obtain $\Gamma_{f,\app} > 1.86$.

\begin{figure}
\hbox{\includegraphics[width=\columnwidth]{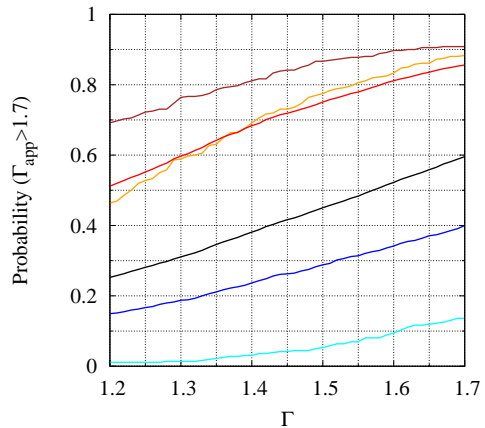}}
\caption{The probability that $\Gamma_{\app}$ will exceed 1.7 is plotted against the underlying $\Gamma$
 when the effect of spiral arms in concentrating the GMCs is considered.
 Simulation parameters are 
$r_\obs=200$ pc, standard velocity distribution, and arm parameters given in text.
The black curve includes all (arm and interarm) 9000 values of $\Gamma_\app$ for all locations.
The blue curve includes the $\Gamma_\app$ values in the strong arm region 
(spheres located in the range $7 < x < 11$ kpc of Figure 7).
The red curve includes the $\Gamma_\app$ values in the interarm region that precedes the strong arm (spheres located in the range $-2 < x < 7$ kpc).
The brown curve includes only the $\Gamma_\app$ values in the sphere centered at $x=6.5$ kpc where the average value of $\Gamma_\app$ reaches its maximum.
The cyan curve includes only the $\Gamma_\app$ values in the sphere centered at $x=10.5$ kpc where the average value of $\Gamma_\app$ reaches its minimum.
The orange curve includes only the $\Gamma_\app$ values in the sphere centered at $x=1.5$ kpc where the Sun is assumed to be located at the final time of the simulation.
}
\label{fig-10} 
\end{figure}

 Finally, we  make a rough estimate the uncertainty in our spiral arm model by presenting the results from simulations that
use other parameters.    We have run our standard spiral arm model but with the \citet{miv17} distribution of GMCs by mass.
Although these two GMC distributions gave very different results for the probability that $\Delta \Gamma_{\app}>0.35$ when
the GMCs were formed at random positions at the solar circle (0.15 for \citet{miv17} versus 0.27 for WM97), when the effect of spiral arms is
included, they differ very little (0.62 versus 0.64) in the interarm region (3).   The reason is that the main effect is that the SFR has been 
suppressed compared to its average solar circle value for roughly 90 Myr, and how the molecular mass is distributed does not matter
significantly.\footnote{ For similar reasons, as discussed above, there is not much dependence on the observing volume.}   We have
also run a case where the fraction of GMC formation in the interarm is $F_I=0.5$ and where $F_S= 0.33$.    These parameters imply that the relation of the star formation rate per unit area in the strong-arm, weak-arm and interarm
regions  is respectively 3:1.5:1.  For $r_{\obs}= 200$ pc and considering only the interarm region described as (3) above, we find that
for this case the probability of $\Delta \Gamma_{\app}>0.35$ is 0.47, as opposed to 0.64 for our standard model.

\section{Summary and Conclusions}

Many observations of young clusters in the Galaxy, as well as global observations of
external galaxies, suggest that the underlying high-mass IMF slope is of order the Salpeter value, $\Gamma=1.35$,
or equivalently that 1.2$\la \Gamma_c \la 1.7$.
However, recent observations 
of stars within $r_\obs\simeq 100$ pc of the Sun \citep{daw10}, $r_\obs\simeq 200$ pc \citep{ryb15},
and $r_\obs\sim 300$~pc \citep{cze14} have led to apparent high-mass slopes
$\Gamma_{f,\app} \simeq 1.7-2.1$.   
The goal of this paper is to reconcile these apparently contradictory observations in the context of an overall
universal IMF.

We have modeled the inhomogeneous distribution of stars at the solar circle caused by the facts that 
(1) they are born in localized GMCs and then randomly move to fill the inter-GMC regions
and (2) that the Sun is in an interarm region of the Galaxy. 
Higher mass stars do not live as long as
solar mass stars. They therefore retain a memory of their birth sites and their surface density distribution is
more inhomogeneous.  In our model, we treat the age-dependent vertical scale heights of stars and 
consider two star formation histories:
the standard history  is that the average star formation rate has declined as exp$(-t/8.3\, \Gyr)$, but we
also consider a constant average star formation rate over the last 3 Gyr.  However, we show that within our simulations
our results are 
very nearly
independent of the models utilized for the SFH or the vertical diffusion.   
We defined $\gapot$ as the apparent value of the high-mass slope after allowing for stellar evolution,
the star formation history, and age-dependent vertical diffusion. For star formation that is spatially and temporally
constant in the plane, this apparent value of the slope is equal to the underlying actual slope. 
Observers generally allow for these effects, so insofar as their models for the SFH and vertical diffusion are
correct, they would find the correct value of $\Gamma$ in this uniform case.
However, clustering
of star formation due to the fact that many stars are born in massive GMCs that are well separated and to the 
effects of spiral arms causes the apparent $\gap$ to differ from $\Gamma$ (for simplicity, we dropped the 
subscript 12 in Section \ref{sec:vert}).
We then defined $\Delta \Gamma_{\app} \equiv
\Gamma_{\app} - \Gamma$
and showed that the distribution of $\Delta \Gamma_{\app}$
that result from sampling various small volumes at the solar circle is independent of the underlying $\Gamma$ if $1.2 < \Gamma < 1.7$.

Our numerical models treat an area of 37.5 kpc$^2$: 25 kpc around the solar circle and 1.5 kpc in radial extent.   In that area,
we form GMCs with their observed mass distribution and at a rate such that the current surface density of GMCs matches 
observation.  We assume that a GMC lasts for 20 Myr.  Inside each GMC,
stars form during the duration of the GMC  and convert 3.8\% of the mass of the cloud to
stars.   
We form stars with mass $m= 1.5-6$ (solar masses)
assuming a power law IMF in this ``high-mass" region that goes as $m^{-\Gamma}$.     We give each star a random velocity
chosen from a velocity dispersion relation observed for these stars  in the plane of the Galaxy at the solar circle and follow their
motion in the Galactic plane.   
The simulation follows the time evolution of 
roughly $2 \times 10^7$ stars  with $m=1.5-6$ for 3 Gyr. The 3 Gyr ensures that we fully sample the PDMF of all the stars in this mass range,
since the lowest mass (i.e., longest lived) star in the simulation ($m=1.5$)
has a main sequence lifetime of 2.6 Gyr.

In each global simulation we stop the evolution after 3 Gyr, and then construct $25$ independent small  (radius $r_{\obs}=100-750$ pc) spheres 
inside the global simulation, and measure the ratio $\calr \equiv \caln_*(1.5<m<3)/\caln_*(3<m<6)$.  
 The ratio is a measure of the PDMF of the stars in this
mass range and inside the given small volume.   We then run $\sim 360$ global simulations, so that we have a large
number (9000)
of small volumes for statistical analysis.   We compare the $\calr $ value obtained in each
volume with the ratio that would be obtained for a given underlying $\Gamma$ and with corrections for
the SFH and the vertical segregation of stars of different age.   As noted above, this comparison
provides an apparent value for the high-mass slope, $\Gamma_\app$. This $\Gamma_\app$ can differ from the underlying global $\Gamma$
 because the  stars in the small observed volume may have
 a deficit or an excess of higher mass stars compared to lower mass stars (i.e., the PDMFs  in small volumes can vary
 from the PDMF taken over the whole solar circle).   
As expected, our models show that the distribution narrows around the underlying $\Gamma$ (i.e., $\Delta \Gamma_{\app}$
goes to zero) as the random velocity dispersion
of the stars increases, as the stellar formation sites are more uniformly distributed, and as the observational volume increases,
since all these effects tend to homogenize the stellar populations.
We present distribution
functions of $\Delta \Gamma_\app$, which show, for example, that for randomly placed observing volumes with radii of 600-100 pc, there is a 
roughly 17-33\% chance that local observations of the solar neighborhood would result in the inference of $\Delta \Gamma_\app \ga
0.35$.  For our standard case $r_{\obs}= 200$ pc we show that the range of GMC mass distributions in the current literature leads to
variations in this chance from 15\% to 27\%.   In other words, there is a moderate probability that the observed $\Gamma_{\app}$ could be $\ga 1.7$ when
the underlying universal $\Gamma = 1.35$.

The value of $\gap$ inferred from observations can also be affected by spiral arms, as noted by \citet{elm06}.
There is strong evidence that 
 the Sun currently lies in an interarm region that is somewhat devoid of recent star formation.
 Using observations that suggest that the SFR in the interarm region is $\sim 30$\% of the total SFR in external
 spiral galaxies, we present curves (Figure 9) that show the probability of $\Gamma_{\app} \ga
 1.7$ as a function of the underlying $\Gamma$ for various samples in arm and interarm regions.
We conclude that the probability of inferring $\Gamma_\app \ga
 1.7$ in the solar neighborhood is 
$\sim 0.6$
for $\Gamma= 1.35$  and for observing radii  $r_{\obs}\la 500$ pc. If the SFR in the interarm region is $\sim 50$\% of the total SFR, this
probability is reduced to $\sim 0.5$.
We therefore propose that the apparent conflict of more global estimates of $\Gamma \sim 1.35$, or $1.2 \la \Gamma \la 1.7$,  with more local observations of $\Gamma_{f,\app}
\simeq 1.7-2.1 $ is caused by the inhomogeneities in the spatial distribution of stars, which increases with stellar mass, 
and by the fact that the Sun lies currently in an interarm region of diminished GMC and star formation.

 We are further motivated  by the data that is becoming available from ESA's Gaia mission, which will enable a determination of both the average PDMF in the Galaxy and its spatial fluctuations.
As we have shown, spatio-temporal variations in the SFR, due to formation of stars in both clusters and in spiral arms,
lead to spatial variations of the PDMF   
that can be directly measured by Gaia.  
 This will provide powerful constraints on
 relative SFR in the arm and interarm regions; as noted in Appendix B, this should become clearer near the corotation radius.
 It will be more challenging to separate variations in the PDMF due to clustering from those due to variations in the IMF,
 but this should be possible using information on stellar velocities, at least for the younger stars.
 Our work is but a first step, showing how clustering affects local PDMFs.

\section*{Acknowledgements}

We acknowledge valuable input from M. Aumer, J. Bland-Hawthorn, C. Dobbs, R. Genzel, K. Masters, M. Miville-Deschenes, E. Ostriker, 
S. Ragan, and D. Soderblom. 
We also thank the referee for a careful reading of the original manuscript and for
numerous suggestions that improved the clarity and organization of the paper.
The research of CFM is supported in part by NASA ATP grant NNX13AB84G.

\

\appendix
\section*{Appendix A. Stellar Velocity Dispersion}

 The stellar velocity dispersion $\sigma(\tau)$ 
increases with the age $\tau$ of the star
due to
gravitational encounters with GMCs.
We consider three possible evolutions of the stellar velocity dispersion with age $\tau$ of the star:
an upper limit, a lower limit, and an intermediate standard expression.
For our upper limit of the velocity dispersion, we adopt the functional form given by \citet{aum16},
\beq
\sigma_{\ABS}(\tau)= \sigma_1 \left( \frac{\tau+\tau_1}{1+\tau_1}\right)^\mu,
\label{eq:sigma}
\eeq
where the age $\tau$ is is expressed in Gyr.  
In Table
3 we give parameter values of Equation (\ref{eq:sigma}) in the radial and azimuthal
directions that fit the observational estimates at the solar circle derived by 
\cite{nor04} and by \citet{deh98} for $0.5 <\tau< 3$ Gyr.
 We adopt this expression for all ages of the stars in the simulation (i.e., $0 <\tau < 2.57$ Gyr, the lifetime
of a $m= 1.5$ star). It is an upper limit because this expression overestimates the dispersion for
$\tau < 0.5$ Gyr as explained below.

\begin{center}
\begin{tabular}{c}
\hspace{-0.5cm}Table 3. $\sigma_{\rm ABS}(\tau)$ parameters
\end{tabular}\\
\hspace{-0.5cm}\begin{tabular}{llcc}
\hline
\hline
& \vline & $R$ &$\phi$ \\
\hline
$\sigma_1$ (km s$^{-1})$& \vline & 22&13\\
$\tau_1$ (Gyr) & \vline& 0.3&0.4\\
$\mu$ & \vline& 0.31& 0.34 \\
\hline
\end{tabular}
\end{center}

This fit to the velocity dispersion is not accurate for ages less than $\sim 500$ Myr (Aumer, private communication);
in particular, it gives velocity dispersions at birth 
[$\sigma_\ABS(0)=(14,\, 8.5)$~km~s\e\
for the radial and azimuthal velocities, respectively]
significantly greater than observed.
\citet{aum16} estimate that due to random motions of the stars at birth and also of the GMCs in which they form,
the initial ($\tau=0$) velocity dispersions in both the radial and azimuthal directions is of order 6 km s$^{-1}$.    
We are mostly interested in the velocities dispersions of the $m\sim 3$ stars, since it is their spatial inhomogeneity that
drives the dispersion in $\Gamma_\app$.   Since they live typically for only $\sim 300$ Myr, their dispersions lie in the
range where    $\sigma_\ABS$ would provide an overestimate.
For our standard expression for the velocity dispersion, 
we assume that the during the first 500 Myr $\sigma$ increases linearly from its initial value to the ABS value as 
\beq
\sigma(\tau)= \left[1-\left(\frac{\tau}{0.5\, \rm{Gyr}}\right)\right] \,6\mbox{ km s\e } +  \left(\frac{\tau}{0.5\, \rm{Gyr}}\right) \sigma_{\ABS}(\tau).
\eeq 
After the first 500 Myr we assume that the velocities  equal  $\sigma_{\ABS}(\tau)$; in particular,
at $\tau=0.5$~Gyr, we have $\sigma_\ABS=(19,\,11)$~km s\e, respectively.
Finally, for the third possible evolution, we adopt
the lower limit of 6 km s$^{-1}$ in each direction for all $\tau$.

To set the mean speed 
in the plane
of a particular star, we pick the speeds $s_x$ and $s_y$ from a Gaussian distribution with standard 
deviations $\langle \sigma_\phi\rangle(\tau)$ and $\langle \sigma_R\rangle(\tau)$, respectively,
where $\langle \sigma\rangle(\tau)$ is the mean velocity dispersion from when the star was born to the end of
the simulation, when the star has an age $\tau$.
With that velocity, we then move  the star from its birth location to its final $x,y$ location at the end of the simulation. 
We assume periodic boundary conditions, so
that, for example, a star that crosses the radial ($y$) boundary is introduced at
the other radial boundary at the same $x$ with the same velocity. 
 Similarly, a star that crosses the azimuthal ($x$) boundary is introduced at
the other azimuthal boundary at the same $y$ with the same velocity.

 The vertical component of velocity sets the scale height of a star of a certain age $\tau$.   Equation (22) in the text describes how we
account for the time-evolving scale height of the stars.

\section*{Appendix B. Spiral Arm Model}

To consider the effect of the spiral arms we assume that the probability of formation of GMCs is enhanced in the arm regions.
 This is the only effect of  
spiral arms in our model: the positions where GMCs form are modified, but not their mass or the global formation rate of GMCs. We consider that the spiral arm 
is a global density wave.
We assume a four-arm model with two strong and two weak arms
(see Section \ref{sec:spiral}), where a fraction $F_{S}$ 
and $F_{W}$ 
of the  GMCs formed in strong and weak arms, respectively.  
The fraction of GMCs formed in the interarm is then
$F_I= 1 -F_S - F_W$.
The value of $F_S+F_W$, the  fraction of stars born in GMCs that were formed 
 in the arms, is uncertain at the solar circle.  We therefore use
observations of the spiral galaxies NGC5194, NGC 628, NGC 6946, M51, 
and M31 made by  \cite{foy10}, \cite{azi11}, and \cite{lee11} to estimate $F_S + F_W=0.7$.   This implies
that $F_I= 0.3$ or that roughly 30\% of the star formation is in interarm regions (these references find a range of
values $F_I\sim 0.1$ to 0.6, so our standard model should be considered exemplary.  We also treat the case where
$F_I=0.5$, $F_S=0.33$ and $F_W= 0.17$.  
In our standard case we adopt $F_I=0.3$,  $F_S=0.5$ and $F_W=0.2$. Gaia should help determine these numbers.

Besides $F_{S}$, $F_{W}$, and the number of arms,
the properties of the spiral arm model are fixed by the pitch angle, $\alpha=12^\circ$ (\cite{val05}),
the full arm width
measured normal to the arm, 
$\Delta _{a}$, and the arm velocity with respect to the LSR, $v_{a}$.  
We adopt $\Delta _{a}=0.8$~kpc from the results summarized by \citet{val14}.
 We do not vary the phase of the spiral structure in our modeling and simply place the Sun in a typical interarm position.
 Note that the 
width of the intersection of the arm and the solar circle is $\Delta_x=\Delta_{a}/\sin\alpha=$3.8 kpc, a few times larger than $\Delta_{a}$. Moreover, 
a GMC that forms within an arm continues to form stars for about 20 Myr, and as a result the $x$-width of enhanced star 
formation region is more than a kpc wider than the enhanced GMC formation region.  
Finally, $v_a=R_0\Omega_p-v_c$, where the angular velocity of the spiral pattern, $\Omega_p$, is estimated
to lie in the range (20 - 25) km s\e\ kpc\e\ \citep{ligerhard16,val17}.  Taking
the average value of $\Omega_p$ and adopting $R_0=8.3$~kpc (rounded from
$8.34\pm0.16$~kpc--\citealp{rei14}) and a circular velocity $v_c=240$ km s\e\ \citet{rei14}, we
find $v_a\simeq -50$~km~s\e.
This implies that it takes 260 Myr to travel from the front of one arm to the
front of the next; the time spent in the interarm region, neglecting velocity changes in the arms, is about 180 Myr.
For these parameter values, the azimuthal ($x$) position of an arm at the top and bottom of the simulation area ($y_T=1.5$
 kpc) differs by
$y_T/\tan\alpha=7.05$ kpc. Since the length of simulation area in the $x$ direction ($x_T=25$ kpc ) is about half of the solar 
circle length, the separation between arms is 
about
$x_T/2$ in a 4-arm model.

 In our simple model the star motions are not modified by their interaction with the spiral structure (except 
implicitly through the dependence of $\sigma$ on $\tau$).  
As seen from the north Galactic pole, the stars and spiral pattern rotate clockwise, but since the
solar circle lies within the corotation radius,
 $R_{\rm CR}=v_c/\Omega_p\simeq 10.5$~kpc.
the arm velocity with respect to the LSR is negative 
in our $x,y$ coordinates.

All the results presented in this paper are for the solar circle. If the observation volumes are 
placed at different Galactocentric radii $R$, the time spent in the interarm region first increases with
 $R$ up to 
the corotation radius
and then decreases.
As consequence, the probability of finding large values of $\Delta\Gamma_{\app}$ will reach a maximum
at the corotation radius. Beyond that 
radius,
the dispersion of $\Delta\Gamma_{\app}$ would decrease with $R$,
as long as the properties of the spiral arms and the stellar diffusion remain about the same.

For a simulation area where $x_T= 25$ kpc and $y_T= 1.5$ kpc, or
$x_T \gg y_T$, the front-side and the back-side of an arm 
(in our model of the Milky Way, this corresponds to the inner and outer sides),
can be described as 
\beq
x_{f}(y,t)=x_{0}-\frac{y}{\tan\alpha} - v_{a}\, (t_0 - t) 
\label{eq:xfront}
\eeq
and
\beq
x_{b}(y,t)=x_{f}(y,t) + \frac{\Delta_{a}}{\sin\alpha},
\label{eq:xback}
\eeq
where $x_{0}=x_{f}(y=0,t=t_0)$ is the final front-side arm position at $t=t_0$ and $y=0$.
In the frame of the simulation area and as seen from the north Galactic pole the arms move to the left
since the Sun is inside corotation ($v_{a}<0$)
and the arm is tilted to the left when $0<\alpha<\pi/2$. 
 Outside corotation, $v_a$ is positive.

The first step in determining the location of a GMC formed at time $t$ is to set
its $y$ position at random in the range 0 to $y_T$.   After that choice, the position of
the front and back side of the arm are determined using Equations (\ref{eq:xfront},\ref{eq:xback}),  and 
finally the $x$ position is picked from the probability distribution $dP(t,y,x)/dx$ in the $x$ range from 0 to $x_T$:
\begin{eqnarray}
\frac{dP(t,x,y)}{dx}&=& \\ 
&&\hspace{-2.2cm}\left\{ \begin{array}{ll}
F_S /\Delta_{x} & {\rm if \,\, ({\it x,y}) \,\, is \,\, within  \,\, a \,\,strong  \,\, arm\,\, at\,\, time\,\, {\it t}}, \\
F_W /\Delta_{x} & {\rm if \,\, ({\it x,y}) \,\, is \,\, within  \,\, a \,\,weak  \,\, arm\,\, at\,\, time\,\, {\it t}}, \\
F_I /(x_T-2\Delta_{x})& {\rm if \,\, ({\it x,y}) \,\, is \,\, in  \,\, the \,\, interarm  \,\, region\,\, at\,\, time\,\, {\it t}}, \\
\end{array}
\right.
\label{eq:prob}
\end{eqnarray}
where $\Delta_{x}= \Delta_{a}/\sin\alpha$.

\bsp	
\label{lastpage}

\begin{thebibliography}{}

\bibitem[Arenou(2011)]{are11} Arenou, F.\ 2011, American Institute of Physics Conference Series, 1346, 107 
\bibitem[Ashworth et al.(2017)]{ash17} Ashworth, G., Fumagalli, M., Krumholz, M.~R., et al.\ 2017, \mnras, 469, 2464
\bibitem[Aumer \& Binney(2009)]{aum09} Aumer, M., \& Binney, J.~J.\ 2009, \mnras, 397, 1286 
\bibitem[Aumer et al.(2016)]{aum16} Aumer, M., Binney, J., \& Sch{\"o}nrich, R.\ 2016, \mnras, 462, 1697 
\bibitem[Azimlu et al.(2011)]{azi11} Azimlu, M., Marciniak, R., \& Barmby, P.\ 2011, \aj, 142, 139 
\bibitem[Bastian et al. (2010)]{bas10} Bastian, N., Covey, K.~R., \& Meyer, M.~R.\ 2010, \araa, 48, 339 
\bibitem[Benjamin(2008)]{ben08} Benjamin, R.~A.\ 2008, Massive Star Formation: Observations Confront Theory, 387, 375
\bibitem[Blaauw(1991)]{bla91} Blaauw, A.\ 1991, NATO Advanced Science Institutes (ASI) Series C, 342, 125
\bibitem[Bressan et al.(2012)]{bre12} Bressan, A., Marigo, P., Girardi, L., et al.\ 2012, \mnras, 427, 127
\bibitem[Cappellari et al.(2012)]{cap12} Cappellari, M., McDermid, R.~M., Alatalo, K., et al.\ 2012, \nat, 484, 485 
\bibitem[Chabrier(2005)]{cha05} Chabrier, G.\ 2005, The Initial Mass Function 50 Years Later, 327, 41 
\bibitem[Chomiuk \& Povich (2011)]{cho11} Chomiuk, L., \& Povich, M.~S.\ 2011, \aj, 142, 197 (CP11)
\bibitem[Churchwell et al.(2009)]{chu09} Churchwell, E., Babler, B.~L., Meade, M.~R., et al.\ 2009, \pasp, 121, 213 
\bibitem[Czekaj et al.(2014)]{cze14} Czekaj, M.~A., Robin, A.~C., Figueras, F., Luri, X., \& Haywood, M.\ 2014, \aap, 564, A102
\bibitem[Dawson \& Schr{\"o}der(2010)]{daw10} Dawson, S.~A., \& Schr{\"o}der, K.-P.\ 2010, \mnras, 404, 917 
\bibitem[Dehnen \& Binney (1998)]{deh98} Dehnen, W., \& Binney, J.\ 1998, \mnras, 294, 429 
\bibitem[Drimmel \& Spergel(2001)]{dri01} Drimmel, R., \& Spergel, D.~N.\ 2001, \apj, 556, 181 
\bibitem[Drimmel et al.(2003)]{dri03} Drimmel, R., Cabrera-Lavers, A., \& López-Corredoira, M. \ 2003, \aap, 409, 205
\bibitem[Elmegreen \& Scalo(2006)] {elm06} Elmegreen, B. G., \& Scalo, J. 2006, \apj, 636, 149
\bibitem[Foyle et al.(2010)]{foy10} Foyle, K., Rix, H.-W., Walter, F., \& Leroy, A.~K.\ 2010, \apj, 725, 534
\bibitem[Georgelin \& Georgelin(1976)]{geo76} Georgelin, Y.~M., \& Georgelin, Y.~P.\ 1976, \aap, 49, 57
\bibitem[Hennebelle \& Falgarone(2012)]{hen12} Hennebelle, P., \& Falgarone, E.\ 2012, \aapr, 20, 55 
\bibitem[Heyer \& Dame(2015)]{hey15} Heyer, M., \& Dame, T.~M.\ 2015, \araa, 53, 583
\bibitem[Hopkins(2012)]{hop12} Hopkins, P.~F.\ 2012, \mnras, 423, 2037 
\bibitem[Just \& Jahrei{\ss}(2010)]{jus10} Just, A., \& Jahrei{\ss}, H.\ 2010, \mnras, 402, 461 
\bibitem[Koo et al.(2017)]{koo17} Koo, B.-C., Park, G., Kim, W.-T., et al.\ 2017, \pasp, 129, 094102
\bibitem[Kroupa et al.(2013)]{kro13} Kroupa, P., Weidner, C., Pflamm-Altenburg, J., et al.\ 2013, Planets, Stars and Stellar Systems.~Volume 5: Galactic Structure and Stellar Populations, 5, 115
\bibitem[Kroupa \& Weidner(2003)]{kro03} Kroupa, P., \& Weidner, C.\ 2003, \apj, 598, 1076 
\bibitem[Lee et al  (2012)]{lee12} Lee, E.~J., Murray, N. \& Rahman, M.\ 2012, \apj,752, 146
\bibitem[Lee et al.(2011)]{lee11} Lee, J.~H., Hwang, N., \& Lee, M.~G.\ 2011, \apj, 735, 75 
\bibitem[Li et al.(2016)]{ligerhard16}Li, Z., Gerhard, O., Shen, J., Portail, M., \& Wegg, C. \ 2016, \apj, 824, 13
\bibitem[Lu et al.(2013)]{ludo13} Lu, J.~R., Do, T., Ghez, A.~M., et al.\ 2013, \apj, 764, 155 
\bibitem[McKee et al.(2015)]{mck15} McKee, C.~F., Parravano, A., \& Hollenbach, D.~J.\ 2015, \apj, 814, 13 
\bibitem[McKee \& Williams(1997)]{mck97} McKee, C.~F., \& Williams, J.~P.\ 1997, \apj, 476, 144 
\bibitem[Miville-Desch{\^e}nes et al.(2017)]{miv17} Miville-Desch{\^e}nes, M.-A., Murray, N., \& Lee, E.~J.\ 2017, \apj, 834, 57 
\bibitem[Muzic et al.(2017)]{muz17} Muzic, K., Schoedel, R., Scholz, A., et al.\ 2017, arXiv:1707.00277 
\bibitem[Nordstr{\"o}m et al.(2004)]{nor04} Nordstr{\"o}m, B., Mayor, M., Andersen, J., et al.\ 2004, \aap, 418, 989 
\bibitem[Offner et al.(2014)]{off14} Offner, S.~S.~R., Clark, P.~C., Hennebelle, P., et al.\ 2014, Protostars and Planets VI, 53
\bibitem[Paresce \& De Marchi(2000)]{par00} Paresce, F., \& De Marchi, G.\ 2000, \apj, 534, 870 
\bibitem[Parravano et al.(2006)]{par06} Parravano, A., McKee, C.~F., \& Hollenbach, D.~J.\ 2006, Revista Mexicana de Fisica Supplement, 52, 1 
\bibitem[Parravano et al.  (2011)]{par11} Parravano, A., McKee, C.~F., \& Hollenbach, D.~J.\ 2011, \apj, 726, 27 (Paper I)	
\bibitem[Prantzos(2008)]{pra08} Prantzos, N.\ 2008, EAS Publications Series, 32, 311 
\bibitem[Reid et al.(2014)]{rei14} Reid, M.~J., Menten, K.~M., Brunthaler, A., et al.\ 2014, \apj, 783, 130 
\bibitem[Renzini(2005)]{ren05} Renzini, A.\ 2005, The Initial Mass Function 50 Years Later, 327, 221  
\bibitem[Rice et al.(2016)]{ric16} Rice, T.~S., Goodman, A.~A., Bergin, E.~A., Beaumont, C., \& Dame, T.~M.\ 2016, \apj, 822, 52 
\bibitem[Robitaille et al.(2012)]{rob12} Robitaille, T.~P., Churchwell, E., Benjamin, R.~A., et al.\ 2012, \aap, 545, A39 
\bibitem[Rosolowsky(2005)]{ros05} Rosolowsky, E.\ 2005, \pasp, 117, 1403 
\bibitem[Rybizki \& Just(2015)]{ryb15} Rybizki, J., \& Just, A.\ 2015, \mnras, 447, 3880 
\bibitem[Scalo(1986)]{sca86} Scalo, J.~M.\ 1986, \fcp, 11, 1 
\bibitem[Scalo(2005)]{sca05} Scalo, J.\ 2005, The Initial Mass Function 50 Years Later, 327, 23  
\bibitem[Schneider et al.(2018)]{schn18} Schneider, F.~R.~N., Sana, H., Evans, C.~J., et al.\ 2018, Science, 359, 69   
\bibitem[Schr{\"o}der \& Pagel(2003)]{schr03} Schr{\"o}der, K.-P., \& Pagel, B.~E.~J.\ 2003, \mnras, 343, 1231
\bibitem[Solomon et al.(1987)]{sol87} Solomon, P.~M., Rivolo, A.~R., Barrett, J., \& Yahil, A.\ 1987, \apj, 319, 730 
\bibitem[Vall{\'e}e(2005)]{val05} Vall{\'e}e, J.~P.\ 2005, \aj, 130, 569 
\bibitem[Vall{\'e}e(2014)]{val14} Vall{\'e}e, J.~P.\ 2014, \aj, 148, 5
\bibitem[Vall{\'e}e(2017)]{val17} Vall{\'e}e, J.~P.\ 2017, \apss, 362, 79 
\bibitem[van Dokkum \& Conroy(2010)]{vandok10} van Dokkum, P.~G., \& Conroy, C.\ 2010, \nat, 468, 940 
\bibitem[Weisz et al.(2015)]{wei15} Weisz, D.~R., Johnson, L.~C., Foreman-Mackey, D., et al.\ 2015, \apj, 806, 198
\bibitem[Williams \& McKee(1997)]{wil97} Williams, J.~P., \& McKee, C.~F.\ 1997, \apj, 476, 166 

\end{thebibliography}
\end{document}